\documentclass[onecolumn]{article}
\usepackage{amsmath,amssymb}
\def\lambdabar{{\mathchar'26\mkern-10mu\lambda}}
\input xy
\xyoption{all}
\xyoption{arc}
\usepackage[curve]{xy}
\xyoption{poly}

\usepackage{hyperref}
\begin{document}
\centerline{\Large A Sketch for a Quantum Theory of Gravity}
\vskip0.2cm
\centerline{April 12, 2005}
\vskip0.3cm
\centerline{\Large Quantum Theory of Galactic Dynamics}
\vskip0.2cm
\centerline{October 21, 2012}
\vskip0.2cm
\centerline{James G. Gilson\quad  j.g.gilson@qmul.ac.uk}
\centerline{School of Mathematical Sciences, Queen Mary} 
\centerline{University of London}
\vskip0.3cm
\centerline {Abstract for Sketch}
\vskip0.3cm
The numerical quantum electronic structure for the energies of the states of the hydrogen like atoms as given by Sommerfeld in 1915-16 is studied and is shown to present a scheme that is able to express a unique {\it observer\/} point of view. The perspective of this observer is essentially how he, if {\it fixed\/} on a trapped electron, would see his and his electron's state of containment within the full atomic quantum state. This particular internal view of a quantum state is then shown to have strong analytic powers in the extremely different numerical scale of gravitation theory.  This unexpected analogy becomes possible when it is recognised that a basic part of gravitation theory can be expressed in terms completely analogous to the quantum relativistic electromagnetic structure involved in Sommerfeld's formula for quantum state energy. The introduction section brings together the essentials for the understanding of the electromagnetic side of this analogy and contains a condensation of some earlier work. The second section {\it Gravitation Analogy\/} develops a theory for gravitation which closely follows the quantum form to give an exact analogy to the Sommerfeld quantum theory of hydrogen atomic states. Quantum Theory of Galactic  Dynamics sections can be found here (\ref{sec-qgme}) onwards.
\section{Introduction}
\setcounter{equation}{0}
\label{sec-intro}
One of the great discoveries of the 20th Century was Sommerfeld's formula for the energy values of the quantum states of the hydrogen
like atoms. The energy values involved were those assumed by a negatively charged electron orbiting a nucleus composed of Z positively charged protons, the charged nucleus being the source of a classical Coulomb electric potential.
In the context of Schr\"odinger wave mechanics which is usually regarded
as non-relativistic \cite{Rin0:2001}, it is surprising that this {\it relativistic\/} formula should appear at all and that its predictions are so
accurate is very remarkable. Sommerfeld's \cite{Som0:1916} formula
for the energy of hydrogen like atoms with a $Z$-valued positive
charge Coulomb central field is,
 \begin{equation}
E_{n,j,Z}=m_0c^2\left(1+\gamma ^2\right)^{-1/2}\label{1.01}
\end{equation}
 where
\begin{equation}
\gamma = \frac{Z\alpha}{
n-j-{1/2}+((j+{1/2})^2-(Z\alpha)^2)^{1/2}} \label{1.02}
\end{equation}
with $n$ what is usually called the principal quantum number and
$j$ with its possible half integral values is the quantum number
for total angular momentum.

The expression (\ref{1.01}) is not
the last word in energy specification for hydrogen like atoms
because it does not include the Lamb shift complications. However,
it is extraordinarily accurate and  shows how the energies depend
on the charge number $Z$. The energies of the first Bohr orbits
for various values of $Z$ are given by $n=1,j=1/2$ and $1\le Z\le
137$ and are
\begin{equation}
E_{1,1/2,Z}=m_0c^2(1-(Z\alpha )^2)^{1/2}. \label{1.03}
\end{equation}
We are particularly interested in the speeds $v_{B,Z}$ with which
the trapped electron moves in these orbits for the various values
of $Z$. The centrifugal equilibrium equation expresses a relation
between the Bohr orbit radius $r_{B,Z}$ and the Bohr orbit
velocity $v_{B,Z}$ and is
\begin{equation}
Ze^2/4\pi\epsilon _0r_{B,Z}=m_0v_{B,Z}^2. \label{1.04}
\end{equation}
Thus (\ref{1.04}) supplies a definition for the velocities in the
first Bohr orbits in terms of the first Bohr radii in the form
\begin{eqnarray}
v_{B,Z}& = &(Ze^2/4\pi\epsilon _0r_{B,Z}m_0)^{1/2}\label{1.05}\\
& = & (Zl_cc^2/r_{B,Z})^{1/2}=Z\alpha c,\label{1.06}
\end{eqnarray}
 the last equality following from the well known result for the circular orbits $r_{B,Z}=r_{B,1}/Z$. Thus in particular,
\begin{equation}
v_{B,1}=\alpha c\label{1.07}
\end{equation}
and
\begin{equation}
v_{B,137}=137\alpha c.\label{1.08}
\end{equation}
Sommerfeld's formula
(\ref{1.01}) for the special case of the first Bohr orbits can easily be derived (\cite{Som0:1916}, \cite{Gil0:2004}, \cite{Gil0:1997})  as will be shown in the next paragraph.

Suppose that a hydrogen-Z atom with its single circling electron
is at rest in some frame of reference $S_0$, the electron being in
the first Bohr orbit. In this frame of reference, the energy
$E(Z)$ of the electron will be given by Sommerfeld's formula  for
a first Bohr orbit. However, for the moment let us simply call the
energy $E(Z)$  and assume we do not know  formula (\ref{1.01}).
Consider the  view of the situation that an observer in  frame of
reference $S$ moving with velocity $v_{B,Z}$ relative to the first
frame such that this velocity is also in the same direction
instantaneously as that of the direction of circling electron. At
the instant of coincidence of these velocities in magnitude and
direction the observer in $S$ will have the electron at rest in
his frame so he will assesses its total energy as being its rest
energy plus its potential energy in the field of the passing
centre of force on the nucleus of the hydrogen-Z atom. This
potential energy is that of one negatively charged electron in the
Coulomb field of the $Z$ positively charged nucleus. This
potential energy is given by $V=-(Ze)e/(4\pi\epsilon  _0r_{B,Z})$.
Thus this observer assesses the total energy $E_T$ of the electron
to be $E_T=m_0c^2-(Ze)e/(4\pi\epsilon _0r_{B,Z})$. This might be
called the electron orientated point  of view. Now consider what
might be called the hydrogen first Bohr state point of view. We
are assuming that the energy of the electron in this state is
given by $E(Z)=M(Z)c^2$, $M(Z)$ being the equivalent mass of this
state. $M(Z)$ is {\it rest mass\/} for this state in the frame $S_0$ in
which the {\it state\/} is at rest. The observer in $S$ will see
this mass as a rest mass $M(Z)$ with a charge $Ze$ moving with
velocity $-v_{B,Z}$. Thus from this point of view he will assess
the energy in motion as $M(Z)c^2/(1-(v_{B,Z}/c)^2)^{1/2}$ and see
it as the nuclear charge $Ze$ moving in the Coulomb potential $V'=
-e(Ze)/(4\pi\epsilon _0r_{B,Z})$ due to his local captured electron.
Thus his assessment of the total energy from the nuclear state in
motion point of view is $E_T'= M(Z)c^2/(1-(v_{B,Z}/c)^2)^{1/2}
-e(Ze)/(4\pi\epsilon _0r_{B,Z})$. The two points of view are two
different ways that the observer in $S$  can perceive the overall
situation. Both of the energies $E_T$ and $E_T'$ are the energies
of  the trapped electron in hydrogen-Z as seen in the frame of
reference $S$ albeit  from the two points of view.  Thus they
should be numerically equal. The two potential energies from the
two points of view are also equal. Thus we have the result
\begin{equation}
\begin{array}{rl}
 m_0c^2 & =  M(Z)c^2/(1-(v_{B,Z}/c)^2)^{1/2}\\
& =  M(Z)c^2/\sin (\chi ^*(Z)). \label{1.09}
\end{array}\end{equation}
$\alpha$ can be given the form
\begin{equation}
\begin{array}{rl}
 \alpha & =  \cos (\chi ^*(Z))/Z,\label{1.10}
\end{array}\end{equation}
so that
\begin{equation}
\begin{array}{rl}
 v_{B,Z} & = Z \alpha  c\\ & = c \cos (\chi ^*(Z)).\label{1.11}
\end{array}\end{equation}
Hence (\ref{1.09}) becomes
\begin{equation}
E(Z)=M(Z)c^2=m_0c^2\sin (\chi ^*(Z)) \label{1.12} \end{equation}
which agrees with (\ref{1.03}).
If the Compton wave length $l_m$ of a rest mass $m$ is defined generally as $\hbar /(m c)$, then (\ref{1.12}) gives
\begin{equation}
l_{m_0}  =  l_{M(Z)}\sin (\chi ^*(Z)). \label{1.13}
\end{equation}
The maximum value for $Z$, $Z=137$, represents choosing the last
member of the hydrogen like set of atoms for analysis, $H_{137}$.
This particular atom is a long way outside present day condition
possibilities. It is a theoretical structure but it has led to
important theoretical discoveries, notably the derivation of a
theoretical formula for $\alpha$
(\cite{Gil0:1999},\cite{Gil0:1991},\cite{Gil0:1994},\cite{Gil0:1996},\cite{BarTip0:1986},\cite{Kli0:1980},
\cite{Kin0:1996},\cite{Bor0:1953},\cite{Wyl0:1969},\cite{CohTay0:1993},\cite{Aoki0:1989},\cite{Sch0:1990}).
Here I shall show that it also has  potential for steering us
towards an understanding of {\it rest mass\/} in terms of
gravitation. This aspect will be pursued in the next section.
However before that, I shall here pull together some earlier
results that will be required in that study. We shall need
definitions for some standard Schr\"odinger quantum theory
quantities. The basic ones are, the classical electron radius
which here will be denoted by $r_e$ but which was denoted in
earlier work by $l_c$, the Compton wavelength of the electron,
$\lambda _C$ and its divided by $2 \pi$ varient, $\lambdabar_C$
here denoted by $2 l_0$. Less basic are the radii of the first
Bohr orbits $r_{B,Z}$ and the velocities in those orbits
$v_{B,Z}$. These definitions are now set out in the order
introduced above together with representations for $\alpha$ and
the general formula for dimensionless coupling constants.
\begin{eqnarray}
   r_e   & = & e^2/(4\pi\epsilon _0 m_0 c^2 )  = l_c \label{1.14} \\
   \lambda _C   & = & h/(m_0 c)  \label{1.15} \\
   \lambdabar _C   & = & \hbar/(m_0 c) = 2l_0 = l_{m_e} \label{1.16} \\
   r_{B,Z}   & = & l_c/(Z \alpha ^2) \label{1.17}  \\
   v_{B,Z}  & = & Z\alpha c \label{1.18} \\
1 & \le Z & \le \ 137 \label{1.19} \\
\alpha & = & r_e/\lambdabar _C=l_c/2l_0=l_{m_e}/r_{B,1}\label{1.20} \\
\alpha   & = & \cos (\chi ^*(Z))/Z\  =  a\ const.\label{1.21} \\
\alpha (n_1,n_2) & = & \frac{n_2\cos(\pi/n_1)\tan(\pi/(n_1n_2))}{\pi}\nonumber\\
&  &                                                   \label{1.22}\\
\alpha & = & \frac{29\cos(\pi/137)\tan(\pi/(29\times 137))}{\pi}\nonumber \\
&   &  \label{1.23}
\end{eqnarray}

In relation to the table above, a remark about the relation between $\alpha (n_1,n_2)$ and the integer parameter $n_1$ will be seen to have important significance in the work to follow. The representation of the fine structure {\it constant\/} $\alpha$ with emphasis on the constant characteristic given by (\ref{1.20}) depends on the many valued parameter $Z$. This representation is very useful in bringing together the physical description for the $137$ members of the hydrogen like family of atoms. As we have seen above it is the member of that set $H_{137}$ with the first Bohr orbit described by the one orbiting electron at the maximum speed $v_{B,137} =  137 \alpha(137,29) c$ that can be used to explain important fundamental physics issues. This speed $v_{B,137}$ is the speed that gives the lowest rest mass $M(137)$ for the rest mass generation process (\ref{1.12}) for the electron rest mass $m_0$. Thus in the quantum electromagnetic context $M(137)$ can be taken to be the most {\it fundamental\/} particle-like mass or smallest mass unit and the smallness of its value depends on the closeness of the light speed multiplier quantity $137\alpha(137,29)$ to unity. In the gravitation analogy context to follow this mass value $M(137)$ will be shown to have an analogy $m_G$ greatly smaller numerically and so much as to be {\it measurably\/} indistinguishable from a photon in a limiting situation.
\newpage
\leftline{Figure 1 Bohr Orbit Hydrogen-Z
Configuration}
\vskip0.5cm
\xy  (3,0)* \xybox{
;<1pc,0pc>:
\POS(4.5,5.5)*+{B_b}
\POS(-.5,5.4)*+{A_a}
\POS(1,4.7)*+{ \pi /Z }
\POS(2,4.5)\ar +(.7,-.7)*+{}
\POS(1.1,3.7)*+{ \chi ^*(Z)}
\POS(2.2,3.4)\ar +(.5,-.5)*+{}
\POS(0,0)*+{O}="a"
\ar +(0,9)*+{A_d}
\ar  +(5.3,7.3)*+{ B_d }
\ar  +(6.06,6.53)*+{*}
\POS(0,8.35)\ar (-3,8.35)*+{c}
\POS(0,0)*+{O}="a"
\POS(-3.28,5.08)*+{a}="c"
,(-3.28,5.08)*{a}="b"
\POS"a" \ar "b"|{r_a(Z)}  \POS"c"
\POS(0,0)*+{O}="a"
\POS(-5.8,2.9)*+{c}="c"
,(-5.8,2.9)*{c}="b"
\POS"a" \ar "b"|{ZR_Q(Z)}  \POS"c"
\POS(0,0)*+{O}="a"
\POS(-6.7,1.4)*+{b}="c"
,(-6.7,1.4)*{b}="b"
\POS"a" \ar "b"|{r_b(Z)}  \POS"c"
\POS(0,0)*+{O}="a"
\POS(-8.6,0)*+{d}="c"
,(-8.6,0)*{d}="b"
\POS"a" \ar "b"|{r_d(Z)}  \POS"c"
\POS(5.1,6.7)*+{}
\ar +(-8.7,0)*+{c'}
\POS(5.8,6.15)*+{}
\ar +(-10.5,0)*+{Z\alpha c}}
*+\xycircle<100pt>{}
*+\xycircle<80pt>{}
*+\xycircle<74pt>{}
*+\xycircle<70pt>{}
*\xybox{
,(-2.5,20)*+{A};(14.7,8.7)*+{B}
**\crv~Lc{~**\dir{}~*{\phantom{\oplus}}
 (2.7,18)&(5,13)&(10,17)&(13,16.7)}}
\endxy
\vskip 0.5cm

Before discussing what this all has to do with gravity let us consider the physical implication of equations (\ref{1.09}) and (\ref{1.12}). Both of these equations represent a simple relation between the rest mass $M(Z)$ of an atomic state and the rest mass $m_0$ of an electron and of course there are many such relations according to the $Z$ value, that is to say, the atomic system chosen.  Equation(\ref{1.09}) clearly means that the electron's rest mass can be fully expressed as the relativistic energy of a {\it particle\/} of rest mass $M(Z)$ in motion with the velocity $v_{B,Z}$. From the (\ref{1.12}) equation, the ratio of the mass $M(Z)$ to the electron's rest mass is in general a relatively small quantity because the orbital velocity $v_{B,Z}$ can have values very close to the velocity of light. The largest value that the orbital velocity $v_{B,Z}$ can have is that which occurs in the case of Hydrogen-137 when $Z=137$ and in this case (\ref{1.12}) gives the ratio of the mass $M(137)$ to the mass $m_0$ of the electron to be the last displayed result below.
\begin{eqnarray}
 v_{B,137}/c & = & 137 \alpha, \label{1.24}\\
 & = & \cos (\chi ^*(137)), \label{1.25}\\
M(137)/m_0 & = & 0.02292\dots\dots\label{1.26}
\end{eqnarray}
Hence equation (\ref{1.09}) essentially expresses the rest mass
$m_0$ of an electron in terms of the sub-mass $M(137)$, about a
200th part of the electron's mass, in motion. Motion here is
generating rest mass by the standard relativistic
kinematics-dynamic process. Thus this analysis gives a model for
the internal dynamics of the electron in terms of sub-particle in
motion structure. This, of course, is the trend of theoretical
physics matter analysis that has occupied most of the last
century, the decomposition of a system into successively smaller
mass sub-systems. The puzzling question is, does this descending
proliferation go on without limit or are there truly fundamental
masses that will ultimately be encountered? However, though that
question cannot now be answered every step downwards is progress
but within the context of Sommerfeld quantum theory we are here at
a base level because $M(137)$ is the smallest valued mass state to
occur. It will be shown that the downward structure picture can be
taken very much further when we examine the gravitation analogy.
There is the well known but curious property   of mass in quantum
systems that very large masses are often very localized and very
small masses can be distributed over large regions. This aspect is
reflected in the way mass appears in the Compton wave lengths such
as $\lambdabar$ for example in equation (\ref{1.15}) where a small
mass in the denominator gives a large wave length and vice versa.
This aspect is related to quantum uncertainty and we shall see
that it has important consequences in the analogy to be developed
in the next section. There is another feature of the rest
mass generation formula (\ref{1.09}) that will be seen to play an
important part in the gravitation story to come. Rest mass is the
most important energy idea that emerged from relativity theory.
Here we can see it has the special role of supplying a {\it
theoretical\/}  observer reference frame from which a particular
energy aspect of the electromagnetic system takes on a very simple
form. Of course, an observer cannot sit on an electron but it
seems that the electron {\it feels\/} its quantum electronic
states relative to its own rest mass platform as though it is
disconnected from such issues as total system rest mass weighting
to the actual space background or to the wider system in which its
states may be imbedded.  The actual mass structure or the centre
of gravity of the massive nucleus of an atom such as that of
$H_{137}$ does not come into the derivation of the relation
between state rest mass $M(Z)$  and kinetically generated  rest
mass $m_0$ of the orbiting electron. If $H_{137}$ were found to
exist physically, besides the $137$ protons in its nucleus, one
might expect it to further contain a large number of neutrons so
that altogether it would be very massive. This all implies the
possible generality of a theoretical {\it unique observer fixed on
the particle\/} viewpoint which is blind to system embedding. The
valuable consequence that follows from this is that we can study
the electron along with its {\it internal\/} states as though it
is in a stand alone condition and just in a low velocity unbound
state. Perhaps this is merely the recognition of a principle that
{\it the internal states of a particle are relative to the
particle's rest frame\/}! I shall refer to this as {\it the
internal relativity principle for a particle\/}. Essentially, this
principle allows role reversal for nucleus and bound rotating
particle. This aspect will also be exploited in the gravitation
context. One may well ask where within the $H_{137}$ atom is the
generating state mass $M(137)$ to be found. The answer to this is
within a transverse speed of light horizon determined by a
rotating radius vector $r_d = c/\omega $ rotating with the
orbiting  electron such that $r_{B,137} \omega = v_{B,137}$. The
values of $r_d$ and $r_{B,137}$ are very close in value,
$r_{B,137}/r_d \approx \cos (\pi /137)\approx 1$. See figure 1,
where the orbiting electron is represented as an extended wave
confined in the angular segment $\pi /137$ between the radii
$r_a(137)$ and $r_b(137)$ its mean position being at radius, $r_c
= r_{B,137}$. The state rest mass $M(137)$ shares the same rest
frame in which the centre of force is at rest.
\section{Gravitation Analogue}
\setcounter{equation}{0}
\label{sec-grava}
The clue that suggests how to quantize gravity comes from one of the three very large pure numbers that have been known about for many years and were extensively studied and used by (Eddington, Dirac, Jordon, Dick, Hayakawa, Carter \cite{Mis0:1973}) and others (\cite{Edd0:1946},\cite{Kil0:1992},\cite{Som0:1916},\cite{Bas0:1996},\cite{BasKil0:1995},\cite{Kil0:1994}) in their theoretical work. Previously, I have denoted this number by $R_{PE}$ indicating that it is the ratio of the electromagnet potential energy of a proton electron pair to the gravitational potential energy of an electron proton pair. This ratio clearly does not depend on the separation of the two particle as it would appear in the numerator and denominator of this ratio and so would cancel. Here I shall denote it by $\xi = R_{PE}$. Its definition is
\begin{eqnarray}
 \xi & = & \alpha \hbar c/(G m_pm_e)\ =\ R_{PE},\label{2.01}
\end{eqnarray}
where $G$ is the constant of gravitation $m_p$ is the rest mass of a  proton and $m_e = m_0$ is the rest mass of an electron. $\alpha$ is the dimensionless coupling constant for the electromagnetic field. Thus we can identify the {\it dimensionless electromagnetic quantum\/} value for gravitational coupling as $ \alpha_G $, say, so that
\begin{eqnarray}
 \alpha_G & = & G m_pm_e /(\hbar c)\label{2.02}
\end{eqnarray} and then
\begin{eqnarray}
 \xi & = & \alpha /\alpha _G. \label{2.03}
\end{eqnarray}

For the purpose of developing this model for a quantum theory of gravity, I shall make the working assumption that all quantum related coupling constants are related to or are members of the set $Q_C = \{ \alpha (n_1, n_2) \}$ defined by equation (\ref{1.22}). As all members of this set are dimensionless, we can then expect to find $\alpha_G$ within this set. Thus I shall represent $\alpha _G$ as
\begin{eqnarray}
 \alpha_G & = & \alpha (N_G^{ }, N_G'),\label{2.04}
\end{eqnarray}
where the pair of integers $N_G, N'_G$
are to be determined. If we compare this equation with equation (\ref{1.12}) we see that the electromagnetic analogue for the maximum value of $N_G$  is $137$ and for $N'_G$ it is $29$. Using equation (\ref{1.22}), the $\xi$ ratio (\ref{2.02}) can be given the more detailed form,
\begin{eqnarray}
 \xi & = & \alpha (137,29) /\alpha (N_G^{ }, N_G'). \label{2.05}
\end{eqnarray}

We can now usefully use a {\it generalization\/} of a formula for
the gravitation constant $G$. The original was suggested by Ross
McPherson \cite{Ros0:2004} based on his observation of a near
numerical coincidence between $\xi$ and the quantum frequency of
the proton's rest mass multiplied by $10^{16}$. The generalized
form to be used here will be expressed in terms of different
physical constants and will have the usual dimensional form
$k_g^{-1}m^3 s^{-2}$ but will be expressed as a function of a time
$t_n$. It has the  definition,
\begin{eqnarray}
G(t_n) & = & \alpha \hbar ^2 H(t_n)/(m_p^2m_e c),\label{2.06}\\
H(t_n) & = & 1/(t_n + t_0),\label{2.07}\\
t_n + t_0 & = & T \ =\ 1.59\times 10^{15}\ s.\label{2.08}
\end{eqnarray}
$H(t_n)$ is Hubble's constant expressed as a function of a time, $t_n$ measured in seconds from a time $t_0 > 0$ near the beginning of the universe. $T$ is a nominal time for which the universe is assumed to have existed. $m_p$ is the rest mass of a proton and $m_e$ is the rest mass of an electron. From  (\ref{2.02}), (\ref{2.06}) and (\ref{2.08}) we can now derive
\begin{eqnarray}
 \alpha _G(t_n) & = & \alpha \hbar /(R(t_n)m_p c) \label{2.09}\\
          & = & r_p /R(t_n). \label{2.10}
\end{eqnarray}
$R(t_n) = cT(t_n)$ is a nominal radius for the universe now and
$r_p$ is the classical radius of the proton. We note that the
dimensionless gravitational coupling constant $\alpha _G(t_n)$
depends on the time $t_n$ through the dependence of the radius of
the universe through the time $t_n$. Equation (\ref{2.09}) enables
us to complete the analogy between quantum theory and gravitation
theory. If we compare equation (\ref{2.10}) with equation
(\ref{1.20}) while assuming that $\alpha _G$ is the gravitational
analogue $\alpha$, it follows that $r_p$ is the gravitational
analogue of $r_e = l_c$ and  a radial length $R_0$ related to
$R(t_n)$ is the gravitational analogue of $\lambdabar _C = 2l_0$.
The rest mass of a graviton $m_G$, say, is naturally defined as
the rest mass which has a Compton wave length $R_0 = \hbar
_G/(m_Gc)$ so that we can write down the gravitational analogue
for equation (\ref{1.12}) as
\begin{equation}
m_G(Z_G)=m_p\sin (\chi _G ^*(Z_G)). \label{2.11}
\end{equation}

Below is set out the list for gravitational quantities analogous to the quantum quantities defined in the list (\ref{1.13})-(\ref{1.22})
\begin{eqnarray}
  r_p   & = & e^2/(4\pi\epsilon _0 m_p c^2 )= \alpha l_{m_p} \label{2.12}\\
 \lambdabar _{G,0}  & = & \hbar_G/(m_G c)\label{2.14} \\
 \hbar _G  & = & \xi \hbar \label{5.14} \\
r_{G,Z_G}   & = & r_p/(Z_G \alpha _G ^2) \label{2.15}  \\
   v_{G,Z_G}  & = & Z_G\alpha_G c\label{2.16}  \\
1 & \le Z_G & \le \ N_G \label{2.17} \\
\alpha_G(t_n) & = & \frac{r_p}{\lambdabar _G} = \frac{r_p}{R(t_n)} = \frac{l_{m_p}}{R(t_n)\alpha ^{-1}} \label{2.18} \\
\alpha _G(t_n)  & = & \cos (\chi^* _G (Z_G))/Z_G  \label{2.19} \\
\alpha _G(n_1,n_2) & = & \frac{n_2\cos(\pi/n_1)\tan(\pi/(n_1n_2))}{\pi}\nonumber\\
&   &  \label{2.20}\\
\alpha _G & = & \frac{N_G'\cos(\pi/N_G)\tan(\pi/(N_G'N^{ }_G))}{\pi}\nonumber \\
&   &  \label{2.21}
\end{eqnarray}

The analogue of the electron in motion on a first Bohr orbit of $H_{137}$ is, what I have indicated above by $G$, a proton in motion on an orbit of radius $r_{G,Z_G}$ with a velocity of $v_{G,Z_G}$ and coupled to a central force via the dimensionless coupling constant $\alpha _G$. Thus the analogue of the relativistic rest mass generation equation (\ref{1.12}) for the proton's rest mass is
\begin{equation}
\begin{array}{rl}
 m_pc^2 & =  m_G(Z_G)c^2/(1-(v_{G,Z_G}/c)^2)^{1/2}\\
& =  m_G(Z_G)c^2/\sin (\chi _G^*(Z_G)). \label{2.22}
\end{array}\end{equation}
$\alpha _G$ can be given the form
\begin{equation}
\begin{array}{rl}
 \alpha _G & =  \cos (\chi _G^*(Z_G))/Z_G.\label{2.23}
\end{array}\end{equation}

The analogy between the structure of the first Bohr orbit of an electron in $H_{137}$ coupled by $\alpha$ to an electromagnetic force centre and the gravitational orbit of a proton coupled by $\alpha _G$ to a gravitational force {\it centre\/} is exact. However, the scales of the analogous systems are greatly different. Below is a list of the approximate numerical values and scale relations for key parameters.
\begin{eqnarray}
\alpha/\alpha _G   & \approx & 2.269 \times 10^{39}\label{2.24} \\
N_G/137 & \approx & 2.2697 \times 10^{39}\label{2.25}\\
M(137)/m_e  & \approx &  0.02292 \label{2.26} \\
m_G/m_e  & \approx &  0.00137568 \label{2.27} \\
M(137)  & \approx &  2.0878919 \times 10^{-32} \label{2.28} \\
m_G  & \approx &  1.2531594783 \times 10^{-33} \label{2.29} \\
m_e  & \approx &  9.10938188 \times 10^{-31} \label{2.30} \\
m_p  & \approx &  1.67262158 \times 10^{-27} \label{2.31} \\
r_{B,137}   & \approx & 3.8626073592 \times 10^{-13} \label{2.32}  \\
r_{G,Z_G}   & \approx & 4.7721275373 \times 10^{23} \label{2.33}  \\
v_{B,137}  & \approx & 2.9971370158 \times 10^8 \label{2.34} \\
v_{G,Z_G}  & \approx & 2.9979245799 \times 10^8 \label{2.35} \\
c  & = & 2.99792458 \times 10^8 \label{2.36}
\end{eqnarray}
\section{Conclusions Sketch}
\setcounter{equation}{0}
\label{sec-conclS}
As suggested in the title, I regard the material presented here as a {\it sketch for a quantum theory of gravitation\/}. However, this material is detailed in that it shows a way in which the very small quantum theory domain of atomic systems can be integrated with the very large cosmology domain within a single theory. The fundamental theories that are called upon to render this construction possible are Sommerfeld's relativistic formula for quantum states together with this author's theory for quantum coupling constants.

A physical result that emerges from this construction is a relativistic dynamical way in which the rest mass of the proton can be seen as formed from the kinematic's of what I have called here, a {\it graviton\/}, a particle with the very small rest mass, $m_G$. As the theory has been developed here, the rest mass of this particle is about a $1/1000$th part of the rest mass of an electron. However, the construction here is limited by experimental knowledge of the true values of the gravitational and other constants and the consequent difficulty of reliably being able to work out the arithmetic of the very large numbers that are involved. It seems that with more reliably accurate information it will be possible to assert that $m_G$ has a {\it limiting\/} value zero. Almost putting it on a par with the photon or neutrino. Further more it's orbits are very large radius circles which are limiting straight lines, if the universe is sufficiently large, again showing a close relation with the photon or neutrino. The angular velocity of the electronic states which are analogous to the $m_G$ states have spin $1/2$ so perhaps it is the neutrino with its spin $1/2$ that is the key to the understanding of quantum gravity.

The theory given here suggests that the rest mass of the proton originates from the graviton kinematics. Possibly this is extendable to some extent to all the hadrons though other considerations may well be involved. Clearly the very complex rest mass relations between the many elementary particles is barely touched by this rest mass generation result for the proton. However it could be a useful start to revelations in that context. It has been shown that an orbiting proton moving on a very large radius path is the gravitational analogue of the orbiting electron in the Sommerfeld quantum electronic theory. Included, is the possibility that the radius of the proton's path is so large that it is indistinguishable from a straight line. The structure given here enables us to identify the gravitational potential analogous to the hydrogen$_{137}$ electric central potential $V(r)=137 \alpha\hbar c/r$ to which a proton is subjected generally and is the kinematic source of its rest mass $m_p$. This is given by
\begin{eqnarray}
 V_G(r_G) & = & N_G\alpha _G \hbar_G c/r_G,\label{3.01}\\
          & = & \xi N_G Gm_pm_e/r_G \label{3.02}.
\end{eqnarray}
From (\ref{3.02}) it can be seen that the central force that holds
the proton in its orbit arises from the gravitational attraction
of a total mass, $ \xi N_Gm_e $, comprised of $N_G$ electronic
sub-masses of strength $\xi m_e$ in analogy with the case of
$H_{137}$ where the central force is composed of a total charge,
$137e$, comprised of $137$ electronic sub-charges of strength $e$.
Thus if we appeal to the particle {\it internal relativity
principle\/} for a proton and its internal gravitational states we
see that the proton's rest mass arises from a centre of force at a
distance $R_G$ from its position. This centre of force is on a
reference frame which will also be host for the internal proton
rest mass gravitational generating mass $m_G$. In keeping with the
usual {\it classical\/} gravitational potential theory,  this
centre of force can be thought off as a mass distribution spread
over a finite sphere within the radius $R_G$. $R_G$ is the radius
of the universe so here we have the situation of greatly distant
masses determining the local rest mass $m_p$ of the proton. This
is a clear indication of the appropriateness of Mach's Principle
(\cite{Gil0:2004}, \cite{Bar0:1995},\cite{Sos0:1980},\cite{Mac0:1893})
in the context of this theoretical construction. This contrasts
sharply with work on Mach's principle in relation to the local
standard of zero rotation (\cite{Gil0:2004}, \cite{Abr0:1990},
\cite{Bel0:1987}, \cite{ Dov0:1980}, \cite{Cav0:1987},
\cite{Gro:1975}, \cite{Ber0:1942}) which does not seem to be
directly related to the local inertia problem. The key result that
connects the quantum domain with the cosmological domain is the
relation between the graviton mass $m_G(t_n)$ and the radius of
the universe $R(t_n)$ at time $t_n$, now. They are both dependent
on time and are related by a rest length radius, $R_0$, of the
universe coinciding with the crossed Compton wavelength $\lambdabar _{G,0}$ of the mass $m_G$ and $\lambdabar _G(t_n) = R(t_n)$.
\begin{eqnarray}
 R(t_n)& = & R_0(1 - (v_{G,Z_G}/c)^2)^{1/2}\label{3.03}\end{eqnarray}

In view of equation (\ref{2.18}), the reader may prefer to define the nominal radius of the universe as the larger value $R^*(t_n) = \alpha ^{-1}R(t_n)$ together with an appropriate $T^*$ and $H^*$. $T^*$ would then have the same order of magnitude as {\it adopted\/} in reference \cite{Mis0:1973} page $738$. There is another significant advantages of adopting the starred definions above because with them there is a starred potential function,
\begin{eqnarray}
 V^*_G(r^*_G) & = & N_G\alpha^{-1}\alpha _G \hbar_G c/r^*_G,\label{3.04}\\
          & = & \xi N_G Gm_p(\alpha ^{-1}m_e)/r^*_G, \label{3.05}
\end{eqnarray}
that replaces the (\ref{3.01},\ref{3.02}) form and
in which the quantity of mass causing the gravitation potential for the very large radius orbit on which the circling proton moves can be seen to be given by $M_U = \xi N_G\alpha ^{-1}m_e$. This turns out to have the same order of magnitude, $10.68\times 10^{53}\ k_g$, as the amount of matter in the universe also as suggested in the reference \cite{Mis0:1973} page $738$. This enhancement option will be considered in an appendix to follow in the next section.

\section{Enhancment Appendix}
\setcounter{equation}{0}
\label{sec-oram}
This section is essentially an appendix containing some elaborations and re-identifications of the age and radius of the universe and other additions. These changes involve a detailed following up of a change in the definition, $T \rightarrow t^*$ of the age of the universe, T as suggested at the end of the previous section.

In earlier sections of this paper, a possible nominal age for the universe was identified from a new formula for the gravitational constant as
\begin{eqnarray}
T & = & \xi\tau _p \label{901}\\
\tau _{m_p}& = & \hbar/(m_pc^2)\label{902}
\end{eqnarray}
where $\tau _{m_p}= \hbar/(m_p c^2) = l_p/c$ is the Compton time interval for a proton.

There is freedom in making this identification because of some dimensionless multipliers appearing in the formula for $G$. Expressed in terms of $T$ or $R$ the formula for $G$ is
\begin{eqnarray}
G & = & \alpha \hbar ^2/(Tm_p^2m_ec) \label{9011}\\
  & = & \alpha \hbar ^2/(Rm_p^2m_e). \label{9021}
\end{eqnarray}
It became apparent that a better correspondence with {\it the age of the universe\/} values suggested from measurement would occur if the $\alpha$ appearing in the numerator of $G$ were incorporated with $T$ to give a transformed value $T^* = \alpha ^{-1} T$. Thus $T^* >T$ and has the same order of magnitude as suggested from the measurement arena. The gravitation constant $G$ in terms of the starred variables becomes
\begin{eqnarray}
G & = & \hbar ^2/(T^*m_p^2m_ec) \label{9012}\\
  & = & \hbar ^2/(R^*m_p^2m_e). \label{9022}
\end{eqnarray}
This redefinition was suggested at the end of the last section and it was
indicated that other quantities would need to be changed in order
for a consistent theory to incorporate the starred definitions. It
turns out that some simple changes have to be made to the quantum
orbital theory to achieve this consistency for the starred
version. The basic change is to replace the classical radius of
the proton, $r_p = \alpha l_p$ in equation (1.17) with its
Compton wave length, $l_p$ as here in equations (4.7), (5.3), (5.4) and (5.5). This
gives a starred version for, $r^*_{G,Z_G}$ for example, replacing
the original $r_{G,Z_G}$ displayed at equation
(2.18) in this case. All the cases are listed below. From an
inspection of the last equality in equation (2.18), it is
then {\it hindsight\/} obvious that this is how it should have
been from the start! Below is a list of the changes to original
variables consequent upon making the change $r_p \rightarrow
l_p $.
\begin{eqnarray}
r_p\ \ \ \ & \rightarrow & l_p\label{903}\\ T\ \ \ \ \ &
\rightarrow & T^* = \alpha ^{-1} T \label{904}\\ R\ \ \ \ \ &
\rightarrow & R^* = \alpha ^{-1} R \label{905}\\ H\ \ \ \ \ &
\rightarrow & H^* = (T^*)^{-1} \label{906}\\ M_U\ \ \   &
\rightarrow & m^*_U = M_U \label{907}\\ \lambdabar_{G,0}\ \  &
\rightarrow & \lambdabar ^*_{G,0}=\alpha ^{-1} \lambdabar_{G,0}
\label{908}\\ r_{G,Z_G} & \rightarrow & r^*_{G,Z_G} =
l_p/(Z_G\alpha ^2_G)\label{909}
\end{eqnarray}

I now wish to make an altogether more subtle change in the {\it interpretational\/} aspects of this theoretical structure. This will involve some additional notation and changes to some physical variables.  Generally the numerical effect of these changes will be negligible and the theory could be left without these additional changes. This would mean that we are taking what might be called a course grained view of the universe and we would not be looking at the detailed structure. However, the enhanced version that emerges is {\it logically\/} more satisfactory and the link and similarity between the small scale quantum aspects and the large scale gravitational aspects becomes more clarified. To see why such changes are desirable we can examine the wave capture diagram fig.1

which originally arose from the electrical-quantum consideration of the orbital structure of $H_{137}$ in relation to the fine structure constant. The most obvious aspect of this diagram are the circular boundaries of radii $r_a$, $r_c$, $r_b$ and $r_d$. There is another  possible boundary missing from the diagram of radius smaller than the shown four that we can call $r_i = r_c \cos (\chi^*(137))$ in the case of $H_{137}$. Thus all the wave structure lies between radii $r_i$ and $r_d$ and the numerical difference between these two radii is $r_d -r_i=r_d (1-\cos ^2(\chi ^*(137))) = r_d\sin ^2 (\chi ^*(137)) \approx r_d (1/137)^2 \approx r_d\times 10^{-4}$. The whole of the geometrical structure between these two extremes would vanish in the thickness of the single circular circumference, if the diagram had not been greatly distorted to make it readable. The philosophy of the work in this article is that the gravitation orbits for a proton, almost straight lines for a free proton, follow the same pattern as in the wave capture diagram with the additional condition that the outer and inner radii for the gravitation orbits differ by vastly smaller amount than the electromagnetic case. This is why I suggest that the new set of changes that are to be implemented are logically and conceptually important but numerically negligible.
However, the thickness of a very thin line on a small disc representing the whole universe could represent $10^8$ human life spans. The two inner radii $r_d$ and $r_b$ are important for QED but need not concern us in the gravitation regime for reasons that will become clear.
\section{Enhancement Variables}
\setcounter{equation}{0}
\label{sec-envar}
The first step in the enhancement is to consider the formulae (2.6) and (2.7) for the gravitation constant $G$ in terms of the time like parameter $T^*$ and abandon the {\it interpretation\/} of $c T^*$ as the radius of the universe. Then reinterpret $R^* = c T^*$ as the nearby quantity the gravitational equivalent of the QED radius of $r_{B,137}$, of first Bohr orbit of $H_{137}$. This equivalent is a quantity, $r^*_{G,N_G}= c t^*_{G,N_G }$, a starred version of what was $r_{G,N_G}$  in $A$ originally. The result of this change is a {\it functional\/} form for $G(r'^*)$, the gravitational constant in terms of $r'^*$.

\begin{eqnarray}
G(r^*_{G,N_G}) & = & \hbar ^2/( t^*_{G,N_G}m_p^2m_e c) \label{9013}\\
  & = & \hbar ^2/( r^*_{G,N_G}m_p^2m_e). \label{9023}
\end{eqnarray}
Equation (\ref{9023}) is to be our fundamental equation connecting the quantum theory for gravitational orbits to the value of the relativistic cosmological quantity $G$. The numerical value of $G$ will now be regarded as {\it defined\/} through equation (\ref{9023}). To show how the radius or the age of the of the universe now comes into the structure we have to examine the other important circular quantum orbits within which the circle of radius $r^*_{G,N_G}$ is to be found, the analogues of $r_i$ and $r_d$ from $H_{137}$ theory. They are defined in ascending size order with $r^*_{G,N_G}$ central in magnitude as
\begin{eqnarray}
r''^* & = & r^*_{G,N_G} \cos (\chi_G)= \frac{ N_G l_p}{ \cos (\chi _G)} \label{9014} \\
r'^* & = & r^*_{G,N_G}= \frac{ N_G l_p}{ \cos^2 (\chi _G)} \label{9024}\\
r^* & = & \frac{r^*_{G,N_G}}{\cos
(\chi_G)}=\frac{ N_Gl_p}{\cos^3(\chi_G)}. \label{9025}
\end{eqnarray}
The notation,
\begin{eqnarray}
l_p & = & \hbar /(m_pc)\label{9026}\\
l'_p & = & \hbar /(m_pc\cos(\chi _G))\label{9027}\\
l''_p & = & \hbar /(m_pc\cos ^2(\chi _G)))\label{9028},
\end{eqnarray} will be used.
The speed on the $r'^*$ is given by $ r'^*\omega  = N_G \alpha_G c$ so that the speed $v^*$ on the $r^*$ orbit will be given by

\begin{eqnarray}
v''^* & = & r''^*\omega = r''^* N_Gc \alpha_G / r'^* = c'' \label{9029}\\
v'^* & = & r'^*\omega = r'^* N_Gc \alpha_G / r'^* = c' \label{9030}\\
v^* & = & r^*\omega = r^* N_Gc \alpha_G / r'^* = c\label{9031}\\
c'\  & = & c\cos (\chi _G)\label{9032}\\
c'' & = & c\cos ^2 (\chi _G).\label{9033}
\end{eqnarray}
Thus we can take these inner and outer boundaries $r''^*$ and $r^*$ within which the main orbit of radius $ r'^* $ lies to represent an annulus fixed to and rotating with the geometrically extended object. The object has a mean position on the radius $ r'^* $. The outer boundary of this annular platform has a transverse velocity of the speed of light and so is naturally to be regarded as the maximum outer boundary for the whole rotating system. This then is good reason to identify $r^*$ as the radius of the universe. This does not mean that the universe is rotating. It rather means that the maximum separation for gravitationally coupled systems coincides numerically with the radius of the universe, $r^*$.

Let us first calculate the orbital angular momentum for an electron in the first Bohr orbit of $H_{137}$. From equations (1.17) and (1.18) this is
\begin{eqnarray}
r_{B,Z} v_{B,Z} m_0 & = & Z\alpha c m_0l_c/(Z\alpha ^2) \label{601}\\
& = & m_0cl_c/\alpha = m_0cl_{m_0} \label{602}\\
& = & \hbar \label{603}
\end{eqnarray}
The orbital angular momentum for the corresponding gravitation orbit of a proton is from equations (\ref{9024}) and (\ref{9030}) and using the starred system
\begin{eqnarray}
r^*_{G,Z_G} v_{G,Z_G} m_p & = & \frac{Z_G\alpha _G c m_pl_p}{(Z_G\alpha^2 _G)} \label{604}\\
& = & \frac{m_pc l_p}{ \alpha _G} \label{605}\\
& = & \frac{\hbar}{\alpha _G} = \frac{N_G\hbar}{\cos (\chi _G(Z_G))}\label{606}
\end{eqnarray}
The starred version of the gravitational potential in which a proton moves has the form on the mean orbit and on the outer boundary as given by
\begin{eqnarray}
V^*_G(r'^*_G) & = & \frac{N_G G m_em_p\alpha ^{-1}_G}{r'^*_G}\label{6070}\\
& = & m_p c'^2\label{6071}\\
V^*_G(r^*_G) & = & \frac{N_G G m_em_p\alpha ^{-1}_G}{r^*_G}\label{6072}\\
& = & m_p c'^2\cos (\alpha _G)\label{6073}\\
r'^*_G & < & r^*_G \label{6074}\\
V^*_G(r'^*_G) & > & V^*_G(r^*_G) \label{6075}
\end{eqnarray}
The {\it gravitation orbit\/} of a proton here means a free proton at rest or in motion under gravitational influence alone. The numerical value of the orbital velocity in the orbit given by the quantum number $Z_G = N_G$ for a given $\alpha _G(N^{}_G,N'_G)$ is the maximum quantum number and so it gives the highest speed in orbit, $ v_{G,N_G}=\cos (\chi^*_G(N_G))c < c$. In general this speed will be very close to but definitely {\it less} than c. Thus when a proton is encountered experimentally apparently not moving at high speed relative to the laboratory it does not mean that such a proton has not got the high velocity in its gravitational orbit being studied here. This is because the observer's laboratory rest frame can in general have a small velocity relative to the proton. In other words, any proton can have a high gravitational velocity  relative to its distant rest mass generating graviton's rest frame.

\section{Numerical Values for Radius, Age and Mass of Universe}
\setcounter{equation}{0}
\label{sec-numvo}
The nominal time for the existence of the universe, its age, will from now on be denoted by $t^*$ and defined by $t^* = r^*_G/c$ where $r^*_G$ is taken to be the current radius of the universe. It is evident that these theoretically obtained values give a very good agreement with the experimentally assessed values and also have the theoretical advantage of greatly clarifying detailed aspects of the structure. The mass of the universe can be obtained by considering the magnitude of the mass that induces the gravitational potential in which the proton moves, or more appropriately one might say, {\it exists\/}. From equation (\ref{6070}) this potential is

\begin{eqnarray}
V^*_G(r'^*_G) & = & \frac{N_G G m_em_p\alpha ^{-1}_G}{r'^*_G}\label{6076}\\
& = & m_p c'^2\label{6077}\\
& = & \frac{GM_um_p}{r'^*_G}\label{6078}\\
M_U & = & N_G m_e\alpha ^{-1}_G\nonumber\\
& = & N_G^2 m_e/\cos (\chi_G(N_G) \label{6079}\\
N_G\  & \approx & 3.11\times 10^{41}\label{6080}\\
\alpha ^{-1}_G & \approx & 3.11\times 10^{41}\label{6081}\\
M_U & \approx & 8.81\times 10^{52} kg\label{6082}\\
M_{U,m} & \approx & 5.68\times 10^{53} kg\label{6083}\\
t^* & \approx & 2.18\times 10^{17} s\label{6084}\\
t^*_m & \approx & 3.1536 \times 10^{17} s\label{6085}
\end{eqnarray}
$M_{U,m}$ is the value for the mass and $t^*_m$ is the age of the universe from observation and measurements as suggested in reference \cite{Mis0:1973} and also admitted as to not being {\it canonical\/}.

The three quantities $r^*_G,\ t^*_G\  and\  M_U$ together with the rest mass of the graviton, $m_G$, all depend on the value of the integer parameter $N_G$ and when I write of {\it theoretical\/} values, I imply that the actual numerical values can only be found if the value of the integer $N_G$ can be input into the calculation. There is no theoretical way to evaluate $N_G$ any more than there is to evaluate $137$ at this time in the physics science story. There is a second parameter $N'_G$ involved but again, unlike in the QED case where it is important, in gravitation theory it does not appear to play a significant part. It can be shown why this is the case as follows.

The QED difference between $\alpha ^{-1}$ and $137$ and the gravitational equivalents have the approximate values
\begin{eqnarray}
\alpha ^{-1}(137,29) - 137 & \approx & 0.036\label{6086}\\
\alpha _G^{-1}(N_G,N'_G) - N_G &  \approx & \frac{\pi ^3}{N^2_G}\left(\frac{1}{2}-\frac{1}{3{N'_G}^2}\right)\nonumber\\
& &\label{6087}\\
&  \rightarrow &\ \frac{\pi ^3}{2 N^2_G} \ \ as\ \ N'_G\rightarrow \infty\nonumber\\
& &\label{6088}\\
\alpha _G^{-1}(10^{42},\infty)- 10^{42} &  \approx &\ 10^{-84}\label{6089}\\
\cos (\chi _G(N_G)) &\rightarrow &1 \ as\ N_G\rightarrow \infty \label{6090}\\
\chi _G(N_G) &\rightarrow &0 \ as\ N_G\rightarrow \infty \label{6091}
\end{eqnarray}

From equations (2.22) and (2.23) we get the relation for the graviton rest mass,
\begin{eqnarray}
m_G & = & m_p (1-(c'/c)^2)^{1/2}\label{704}\\
& = & m_p(1 -\cos ^2(( \chi _G(N_G)))^{1/2}\nonumber\\
& &\label{705}\\
& = & m_p \sin ( \chi _G(N_G))\label{706}\\
& \rightarrow & 0\ as\ N_G\rightarrow \infty\label{707}\\
r^* & = & N_G l_p/\cos ^3 (\chi _G) \label{708}\\
t^* & = & N_G \tau _p/\cos ^3 (\chi _G) \label{709}\\
\tau _p & = & l_p/c\label{710}\\
\cos(\chi _G(N_G))r^* & = & N_Gl''_p.\label{711}
\end{eqnarray}
From equation (\ref{6086}), it can be seen that the inverse fine structure constant differs from the integer $137$ at the second decimal place whereas from equation (\ref{6089}), its gravitational equivalent $\alpha^{-1} _G$ differs from the integer $N_G\approx 10^{42}$ at the $84th$ decimal place, they are essentially equal and from equation (\ref{709}) it can be seen that $t^*$ is essentially equal to $N_G \tau _p$ where $\tau _p$ is a constant eigenfunction {\it time\/} increment associated with the eigenvalue number  $N_G$. Thus the digitally changing quantity $N_G$ can be regarded as a digital parameter determining the age of the universe or we can write $N_G(t^*)$ implying the converse. It makes little difference if you care to regard time as the primary mover for change or regard $N_G$ as determining the time $t^*$ or the age of the universe. Thus in this gravitational theory the quantum integer eigenvalue $N_G$ is of primary importance. The whole story is contained in the concept of {\it projection quantization\/}. This concept is the bridge between quantum theory quantization and relativity theory length contraction. One example is displayed at equation (\ref{711}) for the case of the radius of the universe expressed in terms of the eigen-length $l''_p$. This is just one case among the five others for the quantum orbits in terms of their own $l_p$ versions and the key state integer $N_G$. 
From equations (\ref{6079}) and \ref{709}, it is clear that the amount of mass within this universe is increasing  proportionally to the square of its age. The question then arises as to where this mass is coming from and is conservation of energy being violated in this model? There is a simple answer to this question though it may not be everyone's satisfaction. If we are to talk about expansion at all we do need to have some idea about what the expansion is taking place into. Let us consider the most obvious and simple situation and suppose that the expansion is taking place into a larger containing space and for simplicity assume it is euclidean, $E_{con}$, say. In fact, we have no idea of the actual geometry that might be involved so that this assumption will have to do here. Suppose that our expanding universe is centred at the origin of $E_{con}$. I see no reason why this containing space should be completely empty. It contains an expanding universe but suppose further the centre of that expanding universe is surrounded with a mass distribution that extend in all direction just as a consequence of its containing that universe. At small values for the age of the universe $t^*$, its radius $r^*$ or expanding boundary is within the mass distribution which could itself extend to infinity in $E_{con}$. Let the density of this {\it halo\/} of electronic sized masses be given by, the inverse linear, in $R$ function, $\sigma _G(R)= 3m_e/(\cos (\chi_G(N_G(R/c)) 4 \pi {l_p''}^2 R)$, where $R$ is the distance from the centre of coordinates in $E_{con}$. At time $t^*$ the universe will have expanded to the volume size $ V(r^*)=4 \pi (r^*)^3/3$ so that it will then contain mass of amount $\sigma _G (r^*) V(r^*) = M_U(t^*)$ as given by equation (\ref{6079}). A similar argument could be adduced for more geometrically exotic container spaces. Thus conservation of energy is not threatened. Related material can be found in references, (\cite{Thi0:1958}, \cite{Fey0:1961}, \cite{Bog0:1959}, \cite{Fey20:1961}, \cite{ Dys0:1949}, \cite{Sch0:1953}, \cite{Gra20:1994}, \cite{Cav0:1985}, \cite{Rin20:1961}, \cite{Wei0:1981}, \cite{Wol0:1958}, \cite{Wei0:1988}).
\section{Quantized  Gravitation in the Galactic Context, modelled on the standard Electronic Atomic Case. Section added November 2012}
\setcounter{equation}{0}
\label{sec-qgme}
\vskip0.75cm
\centerline{\Large {\bf Quantum Theory of Galactic Dynamics}}
\centerline{\Large {\bf Cosmological Mass Accumulations}}
\centerline{\Large {\bf Described by s p d f g h i... Symmetry}}
\centerline{\Large {\bf Quantised Gravity and Mass Spectra}}
\centerline{\Large {\bf Within the Lambda Based}}
\centerline{\Large {\bf Dust Universe Model}}
\vskip0.5cm
\centerline{October 21, 2012}
\vskip0.75cm  
\section{Abstract for Galactic Dynamics}
\setcounter{equation}{0}
\label{sec-afgd}
Much of the  introductory section of this paper is devoted to displaying some previously obtained formulae, incorporating a change of notation and variables and giving some explanation of the relation of the work to Newtonian gravitation theory. This section all refers to a quantisation of gravity concentrated on and limited to galaxies with totally spherically symmetric cores  and halos. Only the radial variable $r$ is involved and the emphasis is on the dark matter concept. All the following sections are devoted to generalising the theory to additionally  incorporate a dependence of galactic structure on the $\theta$ and $\phi$ spherical angular coordinates. The theory is derived using Schr\"odinger quantum theory in much the same way as it was used in developing the theory of atomic structure. The theoretical structure to be developed in this papers is a hybrid formulation involving three fundamental theoretical facets, general relativity, Schr\"odinger quantum mechanics and a new theoretical version of isothermal gravity self equilibrium. The combined structure has only become possible because of the discovery of an infinite discrete set of equilibrium states associated with this later theory, the $l$ parameter states. The {\it configuration space\/}  structure of these states has been found to be available in Schr\"odinger theory from a special inverse square law potential which appears to supply an inverse cube self attraction to the origin that maintains galaxies in an isolated steady state self gravity quantum condition. The {\it arbitrary\/} numerical coefficients of these Schr\"odinger states can also depend on $l$ and are appropriately imported from the isothermal equilibrium theory.  The work discussed here is much about how these $l$ states can be interleaved with with the usual Schr\"odinger parameter for angular momentun which I call $l^\prime $ to avoid confusion. The $l$ values have been found to be two possible cases of infinite subsets of the $l^\prime$ values, a $D$ set for the usual mass density distributions in galaxies and an  $P$ set for Einstein's extra pressure term density $3P/c^2$. However these identifications are just a working hypothesis. The usual atomic electron theory approach of separation of variables is used to solve the general gravitational Schr\"odinger equation and it turns out to be rather simpler than the atomic electronic situation. Two version of adapting the Schr\"odinger equation to hold the isothermal $l$ states are given. The first I call a transplant operation that in fact is a replacement of appropriate  Schr\"odinger $l^\prime$ angular momentum state representations with isothermal $l$ state representations. The second version is in the conclusions section and involves simply displaying restricted Schr\"odinger representations that describe various gravitational situations. Also in this section, it is made clear that each of the one component Schr\"odinger representations can be replaced with an equivalent two component representation consisting of a Laplace equation together with a quantised energy equation. Finally, I display the mapping of the angular symmetry defining letters from atomic theory into the quantum theory structure of the isothermal $l$ states. The main products of the theory are a {\it quantisation\/} of the gravitational field with explicitly a refined collections of mass accumulation spectra and a generalisation of Newtonian gravitation theory based on general relativity.         

\vskip 0.2cm 
\centerline{Keywords: Dust Universe, Dark Energy, Dark Matter,}
\centerline{Newton's Gravitation Constant, Einstein's Cosmological Constant,}
\centerline{Cosmological Mass Spectra, Quantised Gravity}
\vskip 0.3cm
\centerline{PACS Nos.: 98.80.-k, 98.80.Es, 98.80.Jk, 98.80.Qc}
\vskip 0.3cm
\section{Introduction Quantum Galactic Dynamics}
\setcounter{equation}{0}
\label{sec-introG}
This paper is a follow up of papers, \cite{71:gil}, \cite{72:gil}, \cite{73:gil}, \cite{75:gil} and \cite{78:gil} of similar titles on the problem of formulating the equation that describes the equilibrium of a gaseous material in a self gravitational equilibrium condition in the galaxy modelling context, \cite{70:wei}, see also, appendix 2 of (\cite{58:gil}). In previous papers I have applied this new theory to examining the rotation curves for galactic star motions. That work established that the velocity curves for these quantized dark matter halos are decisivel flat. That theory also implied a precise formula that can give many possible mass spectra each of which can give a discrete infinity of spectral lines determined by a quantum parameter $l$ with integral values, $1,2,3,\dots,\infty$, starting at unity and extending up to integral $\infty$. Individual spectra are determined by three {\it free\/} parameters, $t_b,r_\epsilon,\beta$. Thus there is a triple continuous infinity number of possibilities when choosing the right spectra for any specific application.
The mass spectra generating function is displayed below
\begin{eqnarray}
& &\quad\quad\quad\quad\quad\quad\quad M_{l+}(r_\epsilon,\beta,t_b) =\nonumber\\
\nonumber\\
& & \frac{c^2\Lambda s(t_b)}{G}\left( \frac{ \beta^{2l}(2l-1)^{4l}2lr_\epsilon^{3-4l}}{3(4l-3)}+  \frac{\beta^{4l-1}(2l-1)^{8l-2}(4l-1) r_\epsilon ^{5-8l} }{(8l-5)}\right).\nonumber\\
\label{l0}
\end{eqnarray}
This can be represented as the sum of two parts arising from a general relativity mass density contribution and the corresponding general relativity pressure distribution as shown below with $D$ and $P$ subscripts consecutively.
\begin{eqnarray}
M_{l+D}(r_\epsilon,\beta,t_b)&=& \frac{c^2\Lambda s(t_b)}{G}\left( \frac{ \beta^{2l}(2l-1)^{4l}2lr_\epsilon^{3-4l}}{3(4l-3)}\right)\label{l1}\\
M_{l+P}(r_\epsilon,\beta,t_b)&=&\frac{c^2\Lambda s(t_b)}{G}\left(\frac{\beta^{4l-1}(2l-1)^{8l-2}(4l-1) r_\epsilon ^{5-8l} }{(8l-5)}\right), \label{l2}
\end{eqnarray}
where $s(t)$ is used to denote the function $\sinh^{-2}((3\Lambda)^{1/2} ct/2)$ from the general relativity dust universe model.
I shall use the rest of this introduction in the form of a subsection to make some changes of notation in the original theory to avoid conflict with the next stage and also improve clarity, before going on to the spherical angles $\theta$ and $\phi$ generalised theory in section (\ref{sec-qsos}).
\subsection{Notational Improvement}
It has turned out that these last two mentioned functions are easier to use and manipulate if the parameter $\beta$ with dimensions, $m^2$, is replaced with a dimensionless parameter $\theta_0$, where
\begin{eqnarray}
 \theta_0=\beta r_e^{-2},\label{l3}
\end{eqnarray}
with $r_\epsilon$ remaining unchanged with the dimension length, $m$.
$\theta_0$ is dimensionless and is identical in meaning to the plain $\theta$ used in the previous paper. I have had to make this last change in order to use plain $\theta$ later, in its usual angular variable context of spherical coordinates.. The discarded symbol $\beta$ will be used later in this paper for a dimensionless version of the radius scalar variable $r$. 
The two functions re-expressed then become 
\begin{eqnarray}
M_{l+D}(r_\epsilon,\theta_0,t_b)&=& \frac{c^2\Lambda s(t_b)}{G}\left( \frac{ (2l-1)^{4l}2l \theta_0^{2 l}r_\epsilon^3
}{3(4l-3)}\right)\label{l4}\\
M_{l+P}(r_\epsilon,\theta_0,t_b)&=&\frac{c^2\Lambda s(t_b)}{G}\left(\frac{ (2l-1)^{8l-2}(4l-1) \theta_0^{4 l-1} r_\epsilon^3
}{(8l-5)}\right), \label{l5}
\end{eqnarray}
If we make the same variable change in the function which gives the total actual mass, $M_l(r)$, associated with a galaxy up to a radius $r$ from its centroid,
\begin{eqnarray}
M_l(r)&=& \frac{c^2\Lambda s(t_b)\beta^{2l}(2l-1)^{4l}}{2G(4l-3)}\left(\frac{4lr_\epsilon^{3-4l}}{3}-r^{3-4l}\right)+\nonumber\\
& & \frac{3c^2\Lambda s(t_b)\beta^{4l-1}(2l-1)^{8l-2}}{2G(8l-5)} \left(\frac{r_\epsilon ^{5-8l}(8l-2)}{3}- r^{5-8l}\right) +\nonumber\\
& &\frac{c^2\Lambda}{3G} r^3,\label{l6}
\end{eqnarray}
we get the replacement
\begin{eqnarray}
M_l(r) &=& \frac{c^2\Lambda s(t_b) \theta_0^{2 l}r_\epsilon^3
 (2l-1)^{4l}}{2 G(4l-3)}\left(\frac{4l}{3}-\left(\frac{r}{r_\epsilon}\right)^{3-4l}\right)+\nonumber\\
& & \frac{3c^2\Lambda s(t_b) \theta_0^{4l-1}r_\epsilon^3
 (2l-1)^{8l-2}}{2 G(8l-5)} \left(\frac{ (8l-2)}{3}- \left(\frac{r}{r_\epsilon}\right)^{5-8l}\right) +\nonumber\\
& & \frac{c^2\Lambda r_\epsilon^3}{3G} \left(\frac{r}{ r_\epsilon}\right)^3.\label{l7}
\end{eqnarray}

I emphasize that the above formula is the {\it total\/} mass as indicated by the plus sign before the dark energy mass, as opposed to {\it total effective\/} mass when there would be a minus sign before the dark energy mass within the galaxy domain. 

\newpage
The Newtonian gravitational potential, $ V_l(r)$, at distance $r$ from the origin produced by this {\it actual\/} total mass is given by
\begin{eqnarray}
V_l(r)&=& \frac{M_l(r)G}{r}=\frac{c^2\Lambda s(t_b) \theta_0^{2 l}r_\epsilon^3
 (2l-1)^{4l}}{2 (4l-3)}\left(\frac{4l}{3r}-\frac{r^{2-4l}}{r_\epsilon^{3-4l}} \right)+\nonumber\\
& & \frac{3c^2\Lambda s(t_b) \theta_0^{4l-1}r_\epsilon^3
 (2l-1)^{8l-2}}{2 (8l-5)} \left(\frac{ (8l-2)}{3r}- \frac{r^{4-8l}}{r_\epsilon^{5-8l}} \right) +\nonumber\\
& & \frac{c^2\Lambda r_\epsilon^3}{3} \left(\frac{r^2}{ r_\epsilon^3}\right).\label{l8}
\end{eqnarray}
The gravitational force produced by this potential per unit mass of subjected particle is given by the gradient, $\hat{\bf r}\cdot\nabla V_l(r)$, of the potential with respect to $r$ in the positive radial direction, $\hat {\bf r}$,
\begin{eqnarray}
\hat {\bf r}\cdot \nabla V_l(r)&=& \frac{c^2\Lambda s(t_b) \theta_0^{2 l}r_\epsilon^3
 (2l-1)^{4l}}{2 (4l-3)}\left(-\frac{4l}{3r^2}+\frac{(4l-2)r^{1-4l}}{r_\epsilon^{3-4l}} \right)+\nonumber\\
& & \frac{3c^2\Lambda s(t_b) \theta_0^{4l-1}r_\epsilon^3
 (2l-1)^{8l-2}}{2 (8l-5)}\left(-\frac{ (8l-2)}{3r^2}+\frac{(8l-4) r^{3-8l}}{r_\epsilon^{5-8l}} \right) +\nonumber\\
& & \frac{c^2\Lambda r_\epsilon^3}{3} \left(\frac{2r}{ r_\epsilon^3}\right).\label{l9}
\end{eqnarray}
\begin{eqnarray}
M_l(r) &=& \frac{c^2\Lambda s(t_b) \theta_0^{2 l}r_\epsilon^3
 (2l-1)^{4l}}{2 G(4l-3)}\left(\frac{4l}{3}-\left(\frac{r}{r_\epsilon}\right)^{3-4l}\right)+\nonumber\\
& & \frac{3c^2\Lambda s(t_b) \theta_0^{4l-1}r_\epsilon^3
 (2l-1)^{8l-2}}{2 G(8l-5)} \left(\frac{ (8l-2)}{3}- \left(\frac{r}{r_\epsilon}\right)^{5-8l}\right) +\nonumber\\
& & \frac{c^2\Lambda r_\epsilon^3}{3G} \left(\frac{r}{ r_\epsilon}\right)^3.\label{l100}
\end{eqnarray}
  
If this is evaluated at $r=r_\epsilon$, the total core mass $ M_l(r_\epsilon)$ is expressed as

\begin{eqnarray}
M_l(r_\epsilon) &=& \frac{c^2\Lambda s(t_b) \theta_0^{2 l}r_\epsilon^3
 (2l-1)^{4l}}{2 G(4l-3) }\left(\frac{4l}{3}-1\right)+\nonumber\\
& & \frac{3c^2\Lambda s(t_b) \theta_0^{4l-1}r_\epsilon^3
 (2l-1)^{8l-2}}{2 G(8l-5) } \left(\frac{ (8l-2)}{3}- 1\right) +\nonumber\\
& & \frac{c^2\Lambda r_\epsilon^3}{3G }.\label{l10}
\end{eqnarray}
\begin{eqnarray}
M_l(r) &=& \frac{c^2\Lambda s(t_b) \theta_0^{2 l}r_\epsilon^3
 (2l-1)^{4l}}{2 G(4l-3)}\left(\frac{4l}{3}-\left(\frac{r}{r_\epsilon}\right)^{3-4l}\right)+\nonumber\\
& & \frac{3c^2\Lambda s(t_b) \theta_0^{4l-1}r_\epsilon^3
 (2l-1)^{8l-2}}{2 G(8l-5)} \left(\frac{ (8l-2)}{3}- \left(\frac{r}{r_\epsilon}\right)^{5-8l}\right) +\nonumber\\
& & \frac{c^2\Lambda r_\epsilon^3}{3G} \left(\frac{r}{ r_\epsilon}\right)^3.\label{l777}
\end{eqnarray}

the ratio, excluding dark energy, of total mass to core mass is given by
\begin{eqnarray}
M_{l,T}(\infty) &=& \frac{2c^2\Lambda s(t_b) \theta_0^{2 l}r_\epsilon^3
 (2l-1)^{4l}l}{3 G(4l-3)}+\nonumber\\
& & \frac{c^2\Lambda s(t_b) \theta_0^{4l-1}r_\epsilon^3
 (2l-1)^{8l-2}(4l-1)}{ G(8l-5)}.\label{l16}
\end{eqnarray}
\begin{eqnarray}
M_{l,C}(r_\epsilon) &=& \frac{c^2\Lambda s(t_b) \theta_0^{2 l}r_\epsilon^3
 (2l-1)^{4l}}{6 G} +\nonumber\\
& & \frac{c^2\Lambda s(t_b) \theta_0^{4l-1}r_\epsilon^3
 (2l-1)^{8l-2}}{2 G}.\label{l17}
\end{eqnarray}
\begin{eqnarray}
n_T(l,\theta)=\frac{M_{l,T}(\infty)}{ m(l,r_\epsilon,t_b,\theta_0)} &=& \frac{4l 
 }{(4l-3)}+\frac{6\theta_0^{2l-1} (2l-1)^{4l-2}(4l-1)}{(8l-5)}\nonumber\\
\label{l21}\\
n_C(l,\theta_0)=\frac{M_{l,C}(r_\epsilon)}{ m(l,r_\epsilon,t_b,\theta_0)} &=& 1 +3\theta_0^{2l-1}(2l-1)^{4l-2}\label{l22}\\
m(l,r_\epsilon,t_b,\theta_0)&=&(2l-1)^{4l}\frac{c^2\Lambda s(t_b) \theta_0^{2 l} r_\epsilon^3}{6 G}.\label{l23}
\end{eqnarray}
Thus the ratio, $R_{T/C}(l,\theta_0)$, of $ M_{l,T}(\infty)$ to $ M_{l,C}(r_\epsilon)$ is given by
\begin{eqnarray}
& &\quad\quad\quad\quad\quad\quad R_{T/C}(l,\theta_0)=\nonumber\\
& &\frac{1}{(1 +3\theta_0^{2l-1}(2l-1)^{4l-2})}\left(\frac{4 l 
 }{(4l-3)} +\frac{6\theta_0^{2l-1} (2l-1)^{4l-2}(4l-1)}{(8l-5)}\right).\nonumber\\
\label{l24}
\end{eqnarray}
The gravitational acceleration field at distance $r$ from the galactic mass centroid is found as follows.
Consider the total actual gravitating mass and its potential
\begin{eqnarray}
V_l(r)=\frac{M_l^\prime (r)G}{r}&=& \frac{M_{l+}^\prime G}{r}+\frac{M_{l-}^\prime G}{r}. \label{k105.8}\\
M_{l-}^\prime(r)&=& A_l\left(  -r^{3-4l}\right)+ B_l \left( -r^{5-8l}\right) \label{k105.9}\\
M_{l+}^\prime (r) &=& M_{l+}(r)  + C_lr^3 .\label{k106} 
\end{eqnarray}
The acceleration per unit mass caused by this potential at distance $r$ from the origin is
\begin{eqnarray}
\hat {\bf r}\cdot \nabla V_l(r)&=& \frac{c^2\Lambda s(t_b) \theta_0^{2 l}r_\epsilon^3
 (2l-1)^{4l}}{2 (4l-3)}\left(-\frac{4l}{3r^2}+\frac{(4l-2)r^{1-4l}}{r_\epsilon^{3-4l}} \right)+\nonumber\\
& & \frac{3c^2\Lambda s(t_b) \theta_0^{4l-1}r_\epsilon^3
 (2l-1)^{8l-2}}{2 (8l-5)}\left(-\frac{ (8l-2)}{3r^2}+\frac{(8l-4) r^{3-8l}}{r_\epsilon^{5-8l}} \right) +\nonumber\\
& & \frac{c^2\Lambda r_\epsilon^3}{3} \left(\frac{2r}{ r_\epsilon^3}\right),\label{k107}
\end{eqnarray}
where the dimensioned parameter $\beta$ has been replaced by the dimensionless parameter $\theta_0 =\beta /r_\epsilon^2$ to clarify the dimensionality of the various contributions. Thus all the last bracketed quantities become dimensionally inverse square but not all variably inverse square. All the coefficients of the large brackets have dimensions $m^3s^{-2}$. Thus all the terms are accelerations. Notably, Newton's gravitation constant $G$ does not occur. In fact, $G$ is replaced by $\Lambda$. This quantized gravitational expression is clearly a substantial generalisation of Newton's law of gravitation. However, we can identify main inverse square law forms as the first terms in the first two large brackets. Both of these terms have minus signs and so represent the usual Newtonian gravitational law of attraction towards the origin. However, both of the large brackets contain also many possible positive signed terms of inverse form determined by the quantum state parameter $l$. They thus represent repulsions from the origin. Clearly the last positive term above represents the repulsive effect of twice Einstein's dark energy term. The two first large brackets originate in the galactic context, from the galactic mass density and the Einstein pressure term mass density from general relativity respectively. 
The inverse repulsive terms in the first two brackets with their positive signs appear to go along with the negative gravity of the last term. They are the terms which simulate negative mass by contributing repulsion and actually exist outside the reference sphere of radius $r$. I mention one more effect from the correction. The negative gravitating term contributed by Einstein's dark energy, the last term above, was left out when I calculated the rotation curve for the small galaxy on the grounds that for a small galaxy it would only make a negligible contribution on account of the smallness of $\Lambda$. However, if is used in such calculations under the corrected version of this theory it would contribute a small positive addition to the rotation curve gradient formula for large galaxies. For sufficiently large galaxies the rotation curves would eventually curve up from their flat condition at very large distances from the origin. There has been mention of observations to this effect. The integer parameter $l$ is closely related to the isotropic index $n$ and was derived in a new version of gravity self equilibrium theory. In the following sections, I shall show how it is related to the quantum angular momentum parameter, also usually denote by $l$ and which to avoid confusion, I shall here denote by $l^\prime$.  Atomic states are usually described, using my changed  notation  by $l^\prime$, and a second Parameter, $m$, sometimes called the $z$ component of angular momentum. With these two angular related parameters the comprehensive theory of atomic state structure has been developed. In the follow pages, I shall show how my new $l$ and the same old $m$ can be used to describe the geometrical state mass distributions of galaxies in much the same way that the angular parameters $l^\prime$ and $m$ are used in the atomic context where it can give considerable graphical and pictorial dressing. Thus I shall take the purely spherically symmetric structure from my theory of galaxies in earlier papers to a much more realistic mathematical representation that can describe many of the possible galactic shapes that are seen.  
\section{Quantum States of Cycling Mass Within a Galaxy}  
\setcounter{equation}{0}
\label{sec-qsos}
In Newtonian mechanics, the energy, $E$, and its equation associated with a particle of mass $\mu$ moving in a potential field $U(r)$ with velocity ${\bf v}$ is given by
\begin{eqnarray}
E&=&\frac{P^2}{2 \mu} +U(r) \label{mm0}\\
P=\left|{\bf P}\right|&=&\mu \left|{\bf v}\right|\label{mm1}
\end{eqnarray}
The essential step in progressing from this so called classical dynamics equation into a quantum mechanics version was the replacement of the two classical dynamical variables energy and momentum, $E$ and  $P$, with operators and necessarily to add to the system something for them to operate on, a state function $\psi$, say.
\begin{eqnarray}
E\rightarrow \hat H&=&i\hbar\frac{\partial}{\partial t} \label{mm2}\\
P_i\rightarrow \hat P_i&=&-i\hbar\frac{\partial}{\partial x_i} \label{mm3}\\
{\bf P}\rightarrow \hat {\bf P}&=&-i\hbar{\bf \nabla} .\label{mm4}\\
{\bf P}^2\rightarrow \hat {\bf P}^2&=&-\hbar^2{\bf \nabla}^2 .\label{mm5}
\end{eqnarray}
Thus in non-relativistic quantum theory the dynamical states $\psi$ of a particle of mass $\mu$ moving in a spherically symmetric field, centred at position $r=0$, can be described by a Schr\"odinger equation, the last equation at, (\ref{m108}) which is just an operator version of (\ref{mm0}) acting on the state function $\psi$. The energy $E$ becomes the Hamiltonian operator, (\ref{m109}),
\begin{eqnarray}
E&=&\frac{P^2}{2 \mu} +U(r) \rightarrow i\hbar\frac{\partial\psi}{\partial t}= -\frac{\hbar^2}{2\mu}\nabla^2\psi + \hat U(r)\psi\label{m108}\\
\hat H&=&i\hbar\frac{\partial}{\partial t} = -\frac{\hbar^2}{2\mu}\nabla^2 + \hat U(r)\label{m109}\\
{\bf \nabla}&=&\frac{{\bf e_1}\ \partial}{\ \ \ \partial r}+\frac{{\bf e_2\ }}{r}\frac{\partial}{\partial \theta}+\frac{{\ \ \ \ \bf e_3}}{r\sin (\theta)}\frac{\partial}{\partial \phi} \label{m109b}\\
\nabla^2&=&\frac{\partial}{r^2\partial r}\left(\frac{r^2\partial}{\partial r}  \right)+ \frac{1}{r^2\sin (\theta)}\left(\frac{\partial}{\partial \theta}\left(\sin (\theta)\frac{\partial}{\partial \theta} \right)+\frac{1}{\sin (\theta)}\frac{\partial^2}{\partial \phi^2}\right)\nonumber\\
 \label{m110}\\
&=& \frac{2\partial}{r\partial r} + \frac{\partial^2}{\partial r^2}+\frac{1}{r^2} \left(\frac{\cot (\theta)\partial}{\partial \theta}+\frac{\partial^2}{\partial \theta^2} +\frac{1}{\sin^2 (\theta)}\frac{\partial^2}{\partial \phi^2}\right) \label{m110a}\\
&=& \frac{2\partial}{r\partial r} + \frac{\partial^2}{\partial r^2}-\frac{\lambda}{r^2},\ say \label{m110aa}\\
{\rm\hat L}&=&-\left(\frac{\cot (\theta)\partial}{\partial \theta}+\frac{\partial^2}{\partial \theta^2} +\frac{1}{\sin^2 (\theta)}\frac{\partial^2}{\partial \phi^2}\right) \label{m110ab}\\
{\rm\hat L}\psi_\lambda &=&\lambda  \psi_\lambda \label{m110ac}\\
x&=&r\sin (\theta)\cos (\phi)\label{m110b}\\
y&=&r\sin (\theta)\sin (\phi)\label{m110c}\\
z&=&r\cos (\theta),\label{m110d}
\end{eqnarray}
if the particle is subjected to the influence of a potential field $\hat U(r)$ when it is at radial distance $r$ from the centre of force. At equation (\ref{mm3}), the momentum vector is shown in component operator form in Cartesian coordinates and in the following line it is shown as a vector operator. However, in discussing systems in three dimensions with a central potential only depending on $r$, they are best represented in terms of spherical polar coordinates, equations (\ref{m110a}) etc. The operator, $\nabla$, is given at equation (\ref{m109b}) and its square is given at equation (\ref{m110}) both in spherical polar coordinates. The square of the momentum operator is obtained from this by multiplication through by $-\hbar^2$.
Under these substitutions of operators replacing  classical physical functions to generate quantum theory, the angular momentum ${\bf L}$ of a particle which was initially defined as the vector product ${\bf r}\wedge {\bf P}$ is converted, using the above transformations, as 
\begin{eqnarray}
{\bf L}={\bf r\wedge {\bf P}}\rightarrow \hat {\bf L}&=&-i\hbar{\bf r}\wedge{\bf \nabla} \label{m111}\\
\hat L_x&=+ &i\hbar(\sin (\phi) \frac{\partial}{\partial \theta }+\cot (\theta)\cos (\phi)\frac{\partial}{\partial\phi})\label{m111a}\\
\hat L_y&=- &i\hbar(\cos (\phi) \frac{\partial}{\partial \theta }-\cot (\theta)\sin (\phi)\frac{\partial}{\partial\phi})\label{m111b}\\
\hat L_z&=& - i\hbar \frac{\partial}{\partial\phi}\label{m111c}\\
{\bf L}^2=({\bf r\wedge {\bf P}})^2\rightarrow \hat {\bf L}^2&=&-\hbar^2({\bf r}\wedge{\bf \nabla})^2\nonumber\\
&=&\frac{-\hbar^2}{\sin (\theta)}\left(  \cos(\theta)\frac{ \partial}{\partial\theta}+\sin (\theta)\frac{\partial^2}{\partial \theta^2} \right)-\nonumber\\
& &\frac{\hbar^2}{\sin^2 (\theta)}\frac{\partial^2}{\partial \phi^2} =\hbar^2\lambda\label{m112}
\end{eqnarray}
and apart from the $-\hbar^2$ appearing here this last expression also importantly appears as the last term divided by $r^2$ in the expression for $\nabla^2$ in the Hamiltonian expression at (\ref{m110}). If we are interested in describing a system in which a particle moves in a planar orbit, There is substantial simplification if we take the direction of the z axis as the same direction as the normal to the plane of motion of the particle which will be the same as the direction of the z-component of angular momentum.
The Schr\"odinger equation at (\ref{m108}) or (\ref{m113}), expanded in terms of these coordinates takes the form (\ref{m114})
\begin{eqnarray}
i\hbar\frac{\partial\psi}{\partial t}= -\frac{\hbar^2}{2\mu}\nabla^2\psi + \hat U(r)\psi .\label{m113}
\end{eqnarray}
\begin{eqnarray}
& &\left( \frac{2\partial}{r\partial r} + \frac{\partial^2}{\partial r^2}+\frac{1}{r^2}\left(\frac{\cot (\theta)\partial}{\partial \theta}+\frac{\partial^2}{\partial \theta^2} +\frac{1}{\sin^2 (\theta)}\frac{\partial^2}{\partial \phi^2}\right)\right)\psi\nonumber\\
  & & \quad\quad\quad\quad\quad\quad\quad\quad +\frac{2\mu}{\hbar^2}( \hat H- \hat U(r))\psi=0 . \label{m114}
\end{eqnarray}
If we can find a solution of the product form%
\begin{eqnarray}
\psi (r,\theta,\phi,t)= {\tt R} (r)\Theta (\theta)\Phi (\phi) e^{-\frac{iE t}{\hbar}} \label{m115}
\end{eqnarray}
in which each variable appears alone in its own function, separation of the variables will have the solved the differential equation for $\psi$. Substituting this $\psi$ into the equation we get
\begin{eqnarray}
& &\left( \frac{2\partial}{r\partial r} + \frac{\partial^2}{\partial r^2}+\frac{1}{r^2} \left(\frac{\cot (\theta)\partial}{\partial \theta}+\frac{\partial^2}{\partial \theta^2} +\frac{1}{\sin^2 (\theta)} \frac{\partial^2}{\partial \phi^2}\right)\right) {\tt R} (r)\Theta (\theta)\Phi (\phi) e^{-\frac{iE t}{\hbar}} \nonumber\\
  & & \quad\quad\quad\quad\quad\quad\quad\quad +\frac{2\mu}{\hbar^2}( i\hbar\frac{\partial}{\partial t}  - \hat U(r)) {\tt R} (r)\Theta (\theta)\Phi (\phi) e^{-\frac{iE t}{\hbar}} =0 . \label{m117}
\end{eqnarray}
If we look at equation (\ref{m117}), we see that there is a possible easy to solve differential equation in $\phi$, if $\frac{\partial^2}{\partial \phi^2}$ is taken to be a constant $-m^2$, say, because this would give, $e^{i\phi}$, or $\sin (\phi)$ and $\cos (\phi)$ solutions. Thus we can write down the differential equation (\ref{m120}) as {\it part of\/} a possible solution to equation (\ref{m117}).
Then taking a further look at equation (\ref{m117}), if we have come across spherical harmonics in terms of $\theta$, we may recognise a second equation in terms of $\theta$ that can be solved if we put
\begin{eqnarray}
 -\lambda=\left(\frac{\cot (\theta)\partial}{\partial \theta}+\frac{\partial^2}{\partial \theta^2} -\frac{m^2}{\sin^2 (\theta)} \right)=-l^\prime(l^\prime +1), \label{m117A}
\end{eqnarray}
giving as a second possible solution equation (\ref{m119}). We can further use this $\lambda$ expression in the original equation to give, a third possibly solvable radial variable equation including the time factor at (\ref{m118}),
\begin{eqnarray}
& &\left( \frac{2\partial}{r\partial r} + \frac{\partial^2}{\partial r^2}-\frac{\lambda}{r^2}+\frac{2\mu}{\hbar^2}( i\hbar\frac{\partial}{\partial t}  - \hat U(r) ) \right) {\tt R} (r) e^{-\frac{iE t}{\hbar}}=0  .\label{m118}
\end{eqnarray}
In this work I use a {\it special\/} form of what is usually called the {\it external\/} potential function $\hat U(r)$ by what in fact is a {\it feed back\/} function of form
\begin{eqnarray}
\hat U(r)= E_{l,m} +\frac{A_l}{r^2}.\label{m11.8}
\end{eqnarray}
The $E_{l,m}$ is a function of $l$ and now including the extra angular related parameter $m$ is  constant in that it does not depend on $r$. The second term is also a constant dependent on $l$ but additionally it appears divided by $r^2$ and so is an inverse square law potential. The $E_{l,m}$ term is a constant total quantity of energy associated with a mass accumulation in  self gravitating isothermal equilibrium. The term involving $r^2$ is a {\it quantum\/} potential that keeps the whole mass assembly in a quantum steady state condition. Thus if this type of potential is used in (\ref{m118}) we get
\begin{eqnarray}
& &\left( \frac{2\partial}{r\partial r} + \frac{\partial^2}{\partial r^2}-\frac{\lambda}{r^2}+\frac{2\mu}{\hbar^2}( i\hbar\frac{\partial}{\partial t}  - E_{l,m} -\frac{A_l}{r^2}) \right) {\tt R} (r) e^{-\frac{iE t}{\hbar}}=0  \label{m118.1}
\end{eqnarray}
Thus now if we use the relation
\begin{eqnarray}
(i\hbar\frac{\partial}{\partial t} -E) e^{-\frac{iE t}{\hbar}}=0 \label{m118.2}
\end{eqnarray}
and identify the energy $E=E_{l,m}$, then we will have separated the radial function with time factor, (\ref{m118.1}), into the two equations (\ref{m118.2} and (\ref{m118.3}).
\begin{eqnarray}
 & &\left( \frac{2\partial}{r\partial r} + \frac{\partial^2}{\partial r^2}-\frac{\lambda}{r^2}-\frac{2\mu}{\hbar^2}\frac{A_l}{r^2} \right) {\tt R} (r) =0   \label{m118.3}
\end{eqnarray}
\begin{eqnarray}
\left(\frac{\partial^2}{\partial\theta^2}+\frac{\cot (\theta)\partial}{\partial \theta}+\lambda -\frac{m^2}{\sin^2(\theta)}\right)\Theta (\theta)=0 \label{m119}
\end{eqnarray}
\begin{eqnarray}
\left(\frac{\partial^2}{\partial\phi^2}+m^2\right)\Phi (\phi)=0 . \label{m120}
\end{eqnarray}
\begin{eqnarray}
\lambda =l^\prime (l^\prime +1)\label{m119b}
\end{eqnarray}
The original differential equation (\ref{m113}) is now expressed in terms of the four equations (\ref{m118.2}), (\ref{m118.3}), (\ref{m119}) and (\ref{m120}) completely separating the variables $t,r,\theta,\phi$. 
The last four just mentioned equations are a parameterisation of equation (\ref{m117}) and whatever  $m$ and $\lambda$ are they can be eliminated between these  equations to give (\ref{m117}) with $E=E_{l,m}$. Thus they are entirely equivalent to the one original equation. We have seen that the last two mentioned equations have known solutions. I have shown in previous papers that the quantization of galactic dynamics require the inverse {\it square\/} law potential introduced above. This will be discussed in the next section.
\section{Quantum Gravity Inverse Square Law Potential}
\setcounter{equation}{0}
\label{sec-qgis}
The mass densities that I am using in developing this theory are tightly defined as arising from Einstein's cosmological constant. All mass density in this theory, positively or negatively gravitating, arises from $\Lambda$. It has recently becoming more apparent that so called {\it dark matter\/} mass is the greatly dominant constituent of most of the positively gravitating mass in the universe. Dark energy on the other hand is {\it positive\/} mass which apparently is negatively gravitating and is present uniformly every where but at extremely low density.
There are two basic mass densities involved in this galaxy modelling project, Einstein's general relativity mass density $ $ and also the additional mass density from his pressure term $3P/c^2$, the pair actually obtained by deriving a polytropic gas equation from a new version formula for describing gravitational self equilibrium. The two densities are respectively at (\ref{m121b}) and (\ref{m123c}),  
\begin{eqnarray}
\grave\rho_{l}(r)&=&\left( \frac{-2 a(2 l-1)^2 }{\pi}\right)^{2l}\left(\frac{r}{r_0}\right)^{-4l}= \theta_0^{2l}(2 l-1)^{4l}\left(\frac{r}{r_\epsilon}\right)^{-4l} \label{m121}\\
\rho_{l}(r)&=& \rho (t_b)\theta_0^{2l}(2 l-1)^{4l}\left(\frac{r}{r_\epsilon}\right)^{-4l} \label{m121b}\\
\grave\rho_{P,l}(r)&=&  3 (\grave\rho_{l} (r))^{\frac{4l-1}{2l}} \label{m122}\\
&=&3   \theta_0^{4l-1}(2 l-1)^{8l-2}\left(\frac{r}{r_\epsilon}\right)^{2-8l}\label{m123}\\
\frac{3 P (r)}{c^2}&=& \rho (t_b) \grave\rho_{P,l}(r)= 3\rho (t_b)\theta_0^{4l-1}(2 l-1)^{8l-2}\left(\frac{r}{r_\epsilon}\right)^{2-8l} \label{m123c}\\
\rho (t) &=&  (3/(8\pi G))(c/R_\Lambda)^2\sinh ^{-2} (3ct/(2 R_\Lambda )). \label{m123d} 
\end{eqnarray}
The same two densities at ( \ref{m121}) and ( \ref{m123}) respectively are dimensionless versions in not having the mass per unit volume pre-multiplier, $\rho(t_b)$ which comes from general relativity. $\rho (t) $ is the general relativity positively gravitating mass density of the substratum at epoch $t$. $\rho (t_b)$ is the time constant density with which the galaxy is born and retains for life. The dimensionless versions are indicated with the top grave accent. If we form two new functions $\psi_{D,l}$ and $\psi_{P,l}$ from (\ref{m121}) and (\ref{m123}) by taking their square roots  and in addition give them a steady state time dependence with a factor $ \exp (-\frac{iE_{D,l}t}{\hbar})$ or $ \exp (-\frac{iE_{P,l}t}{\hbar})$, we obtain
\begin{eqnarray}
\psi_{D,l}(r) &=&\theta_0^{l}(2 l-1)^{2l}\left(\frac{r}{r_\epsilon}\right)^{-2l}\exp (-\frac{iE_{D,l}t}{\hbar}) \label{m124}\\
\psi_{P,l} (r)&=&3 \theta_0^{2l-1/2}(2 l-1)^{4l-1}\left(\frac{r}{r_\epsilon}\right)^{1-4l}\exp (-\frac{iE_{P,l}t}{\hbar}).\label{m125}  
\end{eqnarray}
Mathematically, they can in fact be regarded as the solution of an {\it unusual\/} classical eigen-value problem expressed as follows. Find the {\it eigen-potentials\/} $V_{D,l}({\bf r})$ and $V_{P,l}({\bf r})$   and steady state energy wave functions $ \psi_{D,l}$ and $\psi_{P,l}$ that must be operative if the classical Newtonian energy equation is replaced by what might be called a potential function operator version of Schr\"odinger shape for a system with two distinguishable parts density, $D$ and pressure, $P$.
\begin{eqnarray}
\hat V({\bf r})&=& i\hbar\frac{\partial}{\partial t} +\frac{\hbar ^2}{2\mu} \nabla^{2}\label{m126}\\
\hat V({\bf r})\psi_{D,l} ({\bf r},t))&=& V_{D,l}({\bf r}) \psi_{D,l} ({\bf r},t)\label{m127}\\
\hat V({\bf r})\psi_{P,l} ({\bf r},t))&=& V_{P,l}({\bf r}) \psi_{P,l} ({\bf r},t).\label{m128}
\end{eqnarray}
Thus we have essentially two associated fundamental quantum gravity eigen-potentials. The steady state energies $E_{D,l}$ and  $E_{P,l}$ are obtainable from the formalism and so the eigen-potentials are easily calculated from the last two equations to be 
\begin{eqnarray}
V_{D,l}({\bf r})&=& E_{D,l}+ \frac{\hbar^2 l(2l-1)}{\mu r^2} = E_{D,l}+ \frac{\hbar^2 q_{D,l} }{\mu r^2}\label{m129}\\
V_{P,l}({\bf r})&=& E_{P,l}+ \frac{\hbar^2 (2l-1)(4l-1)}{\mu r^2}= E_{P,l}+ \frac{\hbar^2 q_{P,l} }{\mu r^2}\label{m130}\\
q_{D,l} &=& l(2l-1) \label{m130.1}\\
q_{P,l} &=& (2l-1)(4l-1).\label{m130.2}
\end{eqnarray}
The galactic model discussed earlier had both of these fields present
so that a mass moving under the combination  of these two quantum potential will experience a local {\it quantum\/} gravitational potential which is just their sum as
\begin{eqnarray}
U_{l}({\bf r}) &=& E_{l}+ \frac{\hbar^2 (2l-1)(5l-1)}{\mu r^2} \label{m131}\\
&=& E_{l}+ \frac{\hbar^2 q_l}{\mu r^2}, say \label{m132}\\
q_l&=& (2l-1)(5l-1).\label{m133}
\end{eqnarray}
The densities and potentials just discussed in this section arose in earlier work on this theory in the special case that they had no angular orientation dependence. That is to say they only depended on a radial variable $r$ and so definite {\it pure\/} spherical symmetry only was involved. Clearly this, although being a substantial advance on previous galactic structure theory, is not adequate to describe actual galactic structure which is known to take a variety of non-pure spherical geometric forms just from observation. In this earlier theory, the total masses associated with such density distributions accumulations  were easily calculate just by integrating over $r$. From these integrations the steady state energies $E_{l}$, which only depended on the one quantum parameter $l$, needed to go with the Schr\"odinger description were calculated. In the next section, we shall incorporate angular variation into the structure so that steady state masses or energies will now have to be recalculated and will then depend on the angular structure or angular parameters  of the galactic model. This implies that the mass spectra derived in the earlier work will take on a substantially refined form in then depending on additional angular parameters.   
Let us now return to the radial equation for the spherical quantum state, (\ref{m118}) and check the result of using $V_{D,l}$ to represent the potential $\hat U(r) $,
\begin{eqnarray}
& &\left( \frac{2\partial}{r\partial r} + \frac{\partial^2}{\partial r^2}-
\frac{\lambda}{r^2}+ \frac{2\mu}{\hbar^2}(E - (E_{D,l}+ \frac{\hbar^2 q_{D,l} }{\mu r^2})) \right) {\tt R} (r) e^{-\frac{iE t}{\hbar}}=0.  \label{m134}
\end{eqnarray}
Or when the following condition holds
\begin{eqnarray}
 \lambda&=&l^\prime (l^\prime +1)\label{m135b}
\end{eqnarray}
this becomes
\begin{eqnarray}
& &\left( \frac{2\partial}{r\partial r} + \frac{\partial^2}{\partial r^2}-    
\frac{ l^\prime (l^\prime +1)}{r^2}+ \frac{2\mu}{\hbar^2}(E_{l^\prime}- E_{D,l}- \frac{\hbar^2q_{D,l}}{\mu r^2}) \right) {\tt R} (r) e^{-\frac{iE t}{\hbar}}=0 .\nonumber\\ \label{m136}
\end{eqnarray}
Let us compare equation (\ref{m136} with the radial equation in the case of an electron in orbit under the Coulomb electric potential
\begin{eqnarray}
& &\left( \frac{2\partial}{r\partial r} + \frac{\partial^2}{\partial r^2}-\frac{l^\prime(l^\prime+1)}{r^2}+\frac{2\mu}{\hbar^2}\left(E+\frac{e^2Z}{r}\right) \right) {\tt R} (r)=0 . \label{m137}
\end{eqnarray}
Simply now by putting $ E_{l^\prime} - E_{D,l}=E$ these equations become much alike except for the, {\it oppositely signed\/}, inverse quadratic potential in the one and the inverse linear potential in the other and the additional time factor in the first. Thus, if we can find solutions to equation (\ref{m136}), we will be able to describe massive objects in orbital motion in a galaxy in the same style as electron are quantum mechanically describable in orbital motion in an atom. However, we see also that at this juncture we have parted company with the quantized elctron theory because, we have chosen an extra equation (\ref{m118.2}) that removes the term $E= E_{l^\prime}- E_{D,l}$ and the factor $ e^{-\frac{iE t}{\hbar}}$  from equation (\ref{m136}) in order to achieve a full separation of variables.

 {\Large
\subsection{Radial Equation Solution}
We can now see solving equation (\ref{m136}) as a purely mathematics problem by making it a little simpler and non-dimensional by the dimensionless substitution $\beta$ for the dimensioned quantity $r$. Because the equation to be solved is so much like the hydrogen atom equation, we can follow much the same route as is used to solve the hydrogen quantum radial equation, although a distinct deviation from that route is required. Thus, making the substitutions
\begin{eqnarray}
r&=&\beta \frac{\hbar}{\mu c}\label{m138}\\
E_{l^\prime} - E_{D,l}&=&0,\label{m139}
\end{eqnarray}
where $\frac{\hbar}{\mu c}$ is the Compton wavelength of the mass $\mu$ divided by $2\pi$. The result is
\begin{eqnarray}
& &\left( \frac{2\partial}{ \beta\partial \beta } + \frac{\partial^2}{\partial \beta^2}-    
\frac{ l^\prime (l^\prime +1)}{\beta ^2} -2 \frac{q_{D,l}}{\beta^2} \right) {\tt R} (\beta)=0 .\nonumber\\
 \label{m140}
\end{eqnarray}
}\section{Beyond The Gravitational S State}
\setcounter{equation}{0}
\label{sec-bgss}
Various equations are collected below for ease of reference.
If we inspect equation (\ref{m136}) repeated below at (\ref{m141kc22}), we see two parameter contributions in the inverse $\beta^2$ term, $ l^\prime (l^\prime +1)$ and $2 l(2l-1)$. The first of these terms comes from the $\nabla^2$ of the Schr\"odinger kinetic energy and the second of these terms comes from the external quantum potential energy. If an $s$ state $l^\prime =0$ is chosen, the first of these terms would be zero and only the second term would appear. This would correspond to the earlier work when the case of no dependence on angular variables was involved and it was necessary to introduce the eternal potential involving the $l$ parameter to access the isothermal thermal equilibrium states. It thus becomes apparent that the formula would be exactly the same if the spatially dependant part external potential term had not been introduced but rather the state dependence on the $l^\prime $ parameter was restricted to the form $l^\prime =2 l-1$ to match the $l$ term and where the values of $ l^{\prime}$ there are restricted by  range of value by its $l$ dependence, the same as  in the external potential term. Effectively $l^\prime(l^\prime+1)\rightarrow (2 l -1)2 l$ so matching the $l$ term only the order of this commuting product is changed. Thus we can transplant the isothermal gravitational equilibrium states into the full angular dependent schr\"odinger equation by agreeing  to restrict the angular momentum states of that equation in a specific way to $D$ type quantum gravity states. Thus we start with the following three equations.
\begin{eqnarray}
& &\left( \frac{2\partial}{ \beta\partial \beta } + \frac{\partial^2}{\partial \beta^2}-    
\frac{ l^\prime (l^\prime +1)}{\beta ^2} - \frac{2l(2l-1)}{\beta^2} \right) {\tt R} (\beta)=0 .\nonumber\\
 \label{m141kc22}
\end{eqnarray}
\begin{eqnarray}
\left(\frac{\partial^2}{\partial\theta^2}+\frac{\cot (\theta)\partial}{\partial \theta}+ l^\prime (l^\prime +1)-\frac{m^2}{\sin^2(\theta)}\right)\Theta (\theta)=0 \label{m141kc24}
\end{eqnarray}
\begin{eqnarray}
\left(\frac{\partial^2}{\partial\phi^2}+m^2\right)\Phi (\phi)=0 \label{ m141kc25}
\end{eqnarray}
and bearing in mind the extra parameter $m$, we have
\begin{eqnarray}
V_{D,l,m}({\bf r})&=& E_{D,l,m}+ \frac{\hbar^2 l(2l-1)}{\mu r^2} = E_{D,l,m}+ \frac{\hbar^2 q_{D,l} }{\mu r^2}\label{m141kc28}\\
V_{P,l,m}({\bf r})&=& E_{P,l,m}+ \frac{\hbar^2 (2l-1)(4l-1)}{\mu r^2}= E_{P,l,m}+ \frac{\hbar^2 q_{P,l} }{\mu r^2}\label{m141kc29}\\
q_{D,l} &=& l(2l-1) \label{m141kc30}\\
q_{P,l} &=& (2l-1)(4l-1).\label{m141kc31}
\end{eqnarray}
It is necessary that when the substitution $l^\prime=2 l-1$ into equation (\ref{m141kc22}) is made it is also made into the $\theta$ angle determining equation following, (\ref{m141kc24}). Consequently, I have made the changes in two steps repeating  the relevant three equations at each step so that it is clear that these, rather unusual manipulations, are clearly seen to be mathematically and physically correct.
 
The first step take $l^\prime=2 l-1$ in the relevant equations
\begin{eqnarray}
& &\left( \frac{2\partial}{ \beta\partial \beta } + \frac{\partial^2}{\partial \beta^2}-    
\frac{ 2l(2l -1)}{\beta ^2} - \frac{2l(2l-1)}{\beta^2} \right) {\tt R} (\beta)=0 .\nonumber\\
 \label{m141kc35}
\end{eqnarray}
\begin{eqnarray}
\left(\frac{\partial^2}{\partial\theta^2}+\frac{\cot (\theta)\partial}{\partial \theta}+ (2l-1)2l-\frac{m^2}{\sin^2(\theta)}\right)\Theta (\theta)=0 .\label{m141kc36}
\end{eqnarray}
\begin{eqnarray}
\left(\frac{\partial^2}{\partial\phi^2}+m^2\right)\Phi (\phi)=0 .\label{ m141kc3.6}
\end{eqnarray}
Second step is to remove the spatially determined part of the external potential in equation (\ref{m141kc35}). This step simply changes to zero the term $ \frac{2l(2l-1)}{\beta^2} $ of the external potential because it has becomes effectively part of the normal Sch\"odinger structure. The result is
\begin{eqnarray}
& &\left( \frac{2\partial}{ \beta\partial \beta } + \frac{\partial^2}{\partial \beta^2}-    
\frac{ 2l (2l -1)}{\beta ^2} \right) {\tt R} (\beta)=0 .\nonumber\\
 \label{m141kc39}
\end{eqnarray}
\begin{eqnarray}
\left(\frac{\partial^2}{\partial\theta^2}+\frac{\cot (\theta)\partial}{\partial \theta}+ (2l-1)2l-\frac{m^2}{\sin^2(\theta)}\right)\Theta (\theta)=0 \label{m141kc40}
\end{eqnarray}
\begin{eqnarray}
\left(\frac{\partial^2}{\partial\phi^2}+m^2\right)\Phi (\phi)=0. \label{m141kc41}
\end{eqnarray}
Solutions of equation (\ref{m141kc40}) are the Legendre functions,
\begin{eqnarray}
P_{l^\prime}^m(\cos(\theta))=\sin^m(\theta)T_{l^\prime -m}^m(\cos\theta)), \label{m141kc4.1}
\end{eqnarray}
where the $T$ functions are the tesseral harmonics.
The Legendre functions are finite over the range $0\le \theta\le\pi$ only when $l$ is an integer such that $-(2l-1)=-l^\prime \le m\le +l^\prime =+(2l-1)$ in the $D$ case and $l$ is an integer such that $-(4l-2)=-l^\prime \le m\le +l^\prime =+(4l-2)$ in the $P$ case. In the rotating electron quantum context $m$ is called the magnetic quantum number. In the gravitation context of uncharged rotating mass being discussed in this paper, that name for $m$ is not appropriate.
the  $D$ type field with angular variations is  
\begin{eqnarray}
\psi_D(\beta,\theta,\phi,t)= {\tt R}_{2l-1}(\beta)\Theta_{D,2l-1,m} (\theta)\Phi_m (\phi) e^{-\frac{iE_D t}{\hbar}}. \label{m141kc43}
\end{eqnarray}
We see that the second step makes no difference to the $\theta$ equation and no difference to the $\phi$ equation. The last equation displayed above is the final solution of $D$ type after the changes. The parameter change in the $P$ case is $l^{\prime}\rightarrow 4l-2$ in order for $l^\prime(l^\prime+1)$ to become $(4l-2)(4l-1)=2q_{P,l}$.
\section{Refined Quantum Mass Spectra}
\setcounter{equation}{0}
\label{sec-rqms}
The theory being developed here arises from three distinct areas of study, general relativity, isothermal self-gravity equilibrium and quantum mechanics. The theory leads to a quantum theory of gravity based on Einstein's $\Lambda$, itself involving, the use of a mass spectra system, also based on $\Lambda$, which can supply the source mass accumulations and their distant gravitational influence. The emphasis here and earlier has been on the gravitational field  from galaxies which has been shown to agree with the modern dark matter interpretation of the galactic structure. The mass spectra that have been identified in the structure are, in a sense a side effect of the theory and are strongly dependent on the isothermal self gravity aspect of the structure in that arbitrary coefficients in the quantum structure described in earlier sections have to be imported from self gravity equilibrium theory. Another important feature of the mass densities involved in the theory is that in their raw state they are divergent at the $r=0$ origin of spherical coordinates. To bring this aspect into line with physical reality a minimum value for $r$, $r_\epsilon$, is introduced within in which radius the densities are taken to have the value they have at $r=r_\epsilon$. This section will complete these aspect for the more general case of variable angular dependence that is being developed in this paper.

Suppose that we have obtained the complete form for the now angular dependant quantum theory {\it amplitude} for a possible mass accumulation in terms of all it's parameters as discussed above,  $ \Psi (r,t, l^\prime,\theta,\phi,m)$, say, with only the main parameters being explicitly mentioned.
The amount of accumulated mass that this implies will have the form
\begin{eqnarray}
\int\int\int_{all\ 3\ space}\Psi (r,t, l^\prime,\theta,\phi,m) \Psi^* (r,t, l^\prime,\theta,\phi,m) r^2\sin(\theta ) dr d\theta d\phi, \label{m141kc444}
\end{eqnarray}
where
\begin{eqnarray}
\Psi (r,t, l^\prime,\theta,\phi,m)&=& {\tt R}_{l^\prime} (r)\Theta_{l^\prime,m} (\theta)\Phi_m (\phi)e^{-iE_{l^\prime ,m} t/\hbar}\nonumber\\
&=& {\tt R}_{l^\prime} (r)Y_{l^\prime m}(\theta,\phi) e^{-iE_{l^\prime,m} t/\hbar}\label{m141kc44.4}
\end{eqnarray}
where
\begin{eqnarray}
Y_{l^\prime m}=N_{l^\prime m} e^{i|m|\phi}P_{l^\prime}^m(\cos (\theta))\label{m141kc44.5}
\end{eqnarray}
are the spherical harmonics.
Thus the total accumulated mass, (\ref{m141kc444}), becomes
\begin{eqnarray}
\int_0^\infty ({\tt R}_{ l^\prime} (r))^2r^2dr\int\int_{angles} (Y_{m l^\prime}(\theta,\phi))^2\sin (\theta)d\theta d\phi \label{m141kc445}
\end{eqnarray}
where
\begin{eqnarray}
\int_0^{2\pi}d\phi\int_0^\pi(Y_{m l^\prime}(\theta,\phi))^2 \sin (\theta)d\theta =\frac{4\pi}{\epsilon_m(2 l^\prime+1)}\left(\frac{( l^\prime+m)!}{(l^\prime-m)!}\right) \label{m141kc44}
\end{eqnarray}
and
\begin{eqnarray}
\epsilon_0=1,\  \epsilon_m=2,\  (m=1,2,3,\dots\infty).\label{m141kd44}
\end{eqnarray}
We notice that in the last formula that when $m=0$
\begin{eqnarray}
\int_0^{2\pi}d\phi\int_0^\pi(Y_{0 l^\prime}(\theta,\phi))^2 \sin (\theta)d\theta =\frac{4\pi}{(2 l^\prime+1)} \label{m141kc4.4}
\end{eqnarray}
as $\epsilon_0=1$. However when there is no angular dependence, $l^\prime=0$, the double angular integral has the value $4\pi$, the solid angle subtended by a spherical surface at its centre. In the formation of the original mass spectra functions (\ref{l4}) and (\ref{l5}) this $4\pi$ was included from the start so that to convert those mass quantities into the new angular dependent mass quantities, they need only be multiplied by the function 
\begin{eqnarray}
A(l^\prime,m)=\frac{1}{\epsilon_m(2 l^\prime+1)}\left(\frac{( l^\prime+m)!}{(l^\prime-m)!}\right)\label{l4a}
\end{eqnarray}
with the appropriate value for $l^\prime$ for the $D$ and $P$ cases. In the $D$ case $l^\prime= 2l-1$. In the $P$ case $l^\prime= 4l-2$. The two cases being
\begin{eqnarray}
A(2l-1,m)=\frac{1}{\epsilon_m(4l-1)}\left(\frac{( 2l-1+m)!}{( 2l-1-m)!}\right)\label{l4ab}
\end{eqnarray}
\begin{eqnarray}
A(4l-2,m)=\frac{1}{\epsilon_m(8l-3)}\left(\frac{( 4l-2+m)!}{( 4l-2-m)!}\right).\label{l4ac}
\end{eqnarray}
Thus the mass spectra function for the $D$ and $P$ cases are respectively 
\begin{eqnarray}
M_{l,m,D}(r_\epsilon,\theta_0,t_b)&=& \frac{c^2\Lambda s(t_b)}{G}\left( \frac{ (2l-1)^{4l}2l \theta_0^{2 l}r_\epsilon^3
}{3(4l-3)}\right) A(2l-1,m)\nonumber\\
\label{l4b}\\
M_{l,m,P}(r_\epsilon,\theta_0,t_b)&=&\frac{c^2\Lambda s(t_b)}{G}\left(\frac{ (2l-1)^{8l-2}(4l-1) \theta_0^{4 l-1} r_\epsilon^3
}{(8l-5)}\right)\nonumber\\
& & \quad\quad\quad\quad\quad\quad \times A(4l-2,m).\label{l5b}
\end{eqnarray}
The mass quantity for the amount of mass from $r=0$ up to a radius $r$ including both the mass density distribution and the Einstein pressure distribution for use in forming the total gravitational potential effective at radius $r$ for a galaxy is given by
\begin{eqnarray}
& & M_{l,m}(r) = \frac{c^2\Lambda s(t_b) \theta_0^{2 l}r_\epsilon^3
 (2l-1)^{4l}A(2l-1,m)}{2 G(4l-3)}\left(\frac{4l}{3}-\left(\frac{r}{r_\epsilon}\right)^{3-4l}\right)\nonumber\\
&+& \frac{3c^2\Lambda s(t_b) \theta_0^{4l-1}r_\epsilon^3
 (2l-1)^{8l-2} A(4l-2,m)}{2 G(8l-5)} \left(\frac{ (8l-2)}{3}- \left(\frac{r}{r_\epsilon}\right)^{5-8l}\right).\nonumber\\
\label{l5c}
\end{eqnarray}
The first two of the last three equations can now be seen as a refinement of the original mass spectra formula in that it now depend on the $m$ parameter and from its derivation it is clear that $l$ can now be interpreted as a value associated with a quantised angular momentum of the galaxy. These same remarks apply to the last term which now gives the mass $M_{l,m}(r)$ to be used in calculating the gravitational potential contributed by a not necessarily spherical galaxy at distance $r$ from its centre.

\section{Conclusions galactic Dynamics}
\setcounter{equation}{0}
\label{sec-congd}
This paper fundamentally is about a quantum theory for the dynamics of accumulations of mass that are in a state of self gravitating equilibrium. An obvious class of examples of matter in such a state is the galactic family. The structure of this theory and its mathematics has much in common with the mathematical theory of quantised atomic structure. In this paper, the original theory which only involved {\it spherically symmetric\/} mass distribution, only $r$ dependence, has been substantially extended to include mass distributions that depend on the three variables of three dimensional polar coordinates $r$, $\theta$, and $\phi$ and so making the theory more realistic. However, this is not a theory about the dynamics of arbitrary given quantities of mass. The mass accumulation  quantities involved are themselves derived from  general relativity theory supplemented with  a new theory for self gravitating assembles of mass all based on Einstein's cosmological constant $\Lambda$. Thus there are two immediate results from the theory, mass spectra of the quantised mass distributions and the gravitation equivalent of Newton's inverse square law of gravitation that these quantised masses exert on other distant masses. In previous papers, I have shown that these mass accumulations generate {\it flat velocity-radius\/} curves on neighbouring matter in bound motion so complying with recent observations. In the title of this paper, I have mentioned the atomic quantum states and its s p d f g h i...  classification of atomic states. This classification can be taken over into the quantum theory of galactic mass accumulations using the theory given above. To do this I will first give an alternative version of the {\it transplant\/} version of placing isothermal gravitation equilibrium states into the Schr\"odinger equation given above.
From (\ref{m113}) and (\ref{m114}), we see that the general Schr\"odinger equation operator with its operand $\psi$ can be written in the form
\begin{eqnarray}
\hat S_{l^\prime,m}\psi = i\hbar\frac{\partial\psi}{\partial t}+ \frac{\hbar^2}{2\mu}\nabla^2\psi - \hat U(r)\psi &=&\label{m15D}\\
\frac{\hbar^2}{2\mu}\left( \frac{2\partial}{r\partial r} + \frac{\partial^2}{\partial r^2}+\frac{A_{l^\prime,m} (\theta,\phi )}{r^2} \right)\psi+ ( \hat H- \hat U(r))\psi&=&0\label{m15E}\\
A_{l^\prime,m}(\theta,\phi )\psi =\left(\frac{\cot (\theta)\partial}{\partial \theta}+\frac{\partial^2}{\partial \theta^2} +\frac{B_m(\phi)}{\sin^2\psi (\theta)} \ \right)\psi &=& -l^\prime (l^\prime +1)\psi \label{m15F}\\
B_m(\phi)\psi=\frac{\partial^2\psi}{\partial \phi^2}=-m^2\psi & &\label{m15G}
\end{eqnarray}
provided that $\psi$ is a solution in the separated parameter product form.
It will be useful to express the three dimensional Laplace operator squared, (\ref{m110}), in a similar form as the Schr\"odinger operator above,
\begin{eqnarray}
\hat {\it L}_{l^\prime,m}\psi = \nabla^2\psi &=&\label{m15DD}\\
\left( \frac{2\partial}{r\partial r} + \frac{\partial^2}{\partial r^2}+\frac{A_{l^\prime,m} (\theta,\phi )}{r^2} \right)\psi &=&0\label{m15EE}\\
A_{l^\prime,m}(\theta,\phi )\psi =\left(\frac{\cot (\theta)\partial}{\partial \theta}+\frac{\partial^2}{\partial \theta^2} +\frac{B_m(\phi)}{\sin^2\psi (\theta)} \ \right)\psi &=& -l^\prime (l^\prime +1)\psi \label{m15FF}\\
B_m(\phi)\psi=\frac{\partial^2\psi}{\partial \phi^2}=-m^2\psi & &\label{m15GG}
\end{eqnarray}
The quantum theory of galactic structures has, in earlier papers and in this paper, identified four {\it fundamental\/} types of solution each with its own Schr\"odinger equation $V({\bf r})$ as follows
\begin{eqnarray}
V_{Ds,l}({\bf r})&=& E_{Ds,l}+ \frac{\hbar^2 l(2l-1)}{\mu r^2} = E_{Ds,l}+ \frac{\hbar^2 q_{Ds,l} }{\mu r^2}\label{ m15H}\\
V_{Ps,l}({\bf r})&=& E_{Ps,l}+ \frac{\hbar^2 (2l-1)(4l-1)}{\mu r^2}= E_{Ps,l}+ \frac{\hbar^2 q_{Ps,l} }{\mu r^2}\label{m15I}\\
V_{D,l,m}({\bf r})&=& E_{D,l,m}\label{ m15J}\\
V_{P,l,m}({\bf r})&=& E_{P,l,m}. \label{m15K}
\end{eqnarray}
From (\ref{m15D}), the four Schr\"odinger equation operators that go with these potentials are respectively
\begin{eqnarray}
\hat S_{0,0,D}\psi &=& i\hbar\frac{\partial\psi}{\partial t}+ \frac{\hbar^2}{2\mu}\nabla^2\psi - V_{Ds,l}({\bf r})\psi= \frac{\hbar^2}{2\mu}\nabla^2\psi - \frac{\hbar^2 q_{Ds,l} }{\mu r^2}\psi\nonumber\\
&=&\frac{\hbar^2}{2\mu}\hat {\it L}_{2l-1,0}\psi ,\quad\quad i\hbar\frac{\partial}{\partial t}\psi = E_{D,l}\psi  \label{m15L}\\
\hat S_{0,0,P}\psi &=& i\hbar\frac{\partial\psi}{\partial t}+ \frac{\hbar^2}{2\mu}\nabla^2\psi - V_{Ps,l}({\bf r})\psi=\frac{\hbar^2}{2\mu}\nabla^2\psi - \frac{\hbar^2 q_{Ps,l} }{\mu r^2}\psi\nonumber\\
&=&\frac{\hbar^2}{2\mu}\hat {\it L}_{4l-2,0}\psi ,\quad\quad i\hbar\frac{\partial}{\partial t}\psi = E_{P,l}\psi  \label{m15M}\\
\hat S_{2l-1,m,D}\psi &=& i\hbar\frac{\partial\psi}{\partial t}+ \frac{\hbar^2}{2\mu}\nabla^2\psi - V_{D,l}({\bf r})\psi,\ V_{D,l}({\bf r})= E_{D,l,m}\nonumber\\
&=&\frac{\hbar^2}{2\mu}\hat {\it L}_{2l-1,m}\psi ,\quad\quad i\hbar\frac{\partial}{\partial t}\psi = E_{D,l,m}\psi  \label{m15N}\\
\hat S_{4l-2,m,P}\psi &=& i\hbar\frac{\partial\psi}{\partial t}+ \frac{\hbar^2}{2\mu}\nabla^2\psi - V_{P,l}({\bf r})\psi,\ V_{P,l}({\bf r})= E_{P,l,m}\nonumber\\
&=&\frac{\hbar^2}{2\mu}\hat {\it L}_{4L-2,m}\psi ,\quad\quad i\hbar\frac{\partial}{\partial t}\psi = E_{P,l,m}\psi  . \label{m15O}
\end{eqnarray}
These are the basic Schr\"odinger equations for the $D$ and $P$ type solutions. More complex solutions such as the galactic solutions which involve simultaneously a sum of $D$ type for Einstein's mass density term and a $P$ for his pressure term come from superposition of solutions. In the list of Schr\"odinger operators and operands above each Schr\"odinger case is followed, one step down, by two equations, an $l^\prime$ restricted Laplace operator with operand  and a quantised energy equation which together are equivalent to the preceding Schr\"odinger equation.

 The first two cases above are from the pre-angular dependent work only involve $s$ states as indicated by the $s$ and $l^\prime=0$ subscript on the Schr\"odinger operator. However, for these first two cases, all the isothermal gravitational equilibrium states map into an $s$ state giving the mass spectra found in earlier work, (\ref{l4}) and (\ref{l5}). The first two cases above look very much like the second two cases above. However, for the first two cases the angular momentum content of the Laplace versions of these two cases are picked up from the {\it non-angular\/} related external potential term in $l$ whereas, the angular momentum content of the Laplace equation for the second two cases is picked up from the {\it angular\/} content of the $l^\prime$ term at (\ref{m15F}).  The last two cases above $D$ and $P$ map the quantum lettered states, $l^\prime=2l -1$, and $l^\prime=4l-2$ respectively into values of the isothermal gravitational equilibrium states $(l:1,2,3,4,\dots)$ as follows;\quad
For $D$ states $l^\prime=2l -1$  
\begin{eqnarray}
l&=&1\rightarrow l^\prime =1\equiv p\label{m151}\\
l&=&2\rightarrow l^\prime =3\equiv f\label{m152}\\
l&=&3\rightarrow l^\prime =5\equiv h \label{m153}\\
l&=&4\rightarrow l^\prime =7\equiv k \label{m154}\\
l&=&5\rightarrow l^\prime =9\equiv m \label{m155}\\
\vdots &=&\quad\quad\quad\dots\label{m156}
\end{eqnarray}
For $P$ states $ l^\prime=4l -2$  
\begin{eqnarray}
l&=&1\rightarrow l^\prime =2\equiv d\label{m157}\\
l&=&2\rightarrow l^\prime =6\equiv i\label{m158}\\
l&=&3\rightarrow l^\prime =10\equiv m \label{m159}\\
l&=&4\rightarrow l^\prime =14\equiv t \label{m160}\\
l&=&5\rightarrow l^\prime =18\equiv x \label{m161}\\
\vdots &=&\quad\quad\quad\dots\quad\quad\quad .\label{m162}
\end{eqnarray}
It is noticeable that with advancing $l$ values of the isothermal equilibrium states  the $P$ solutions advance along the letter classification of the angularly involved states much faster than do the $D$ solutions.
There may well be other and better ways to get and express the results of this paper. The course I have chosen is, I think, likely to be the simplest. The reason for emphasising and using the {\it transplant\/} idea is that it seems to expose something very fundamental about what I have identified as the inverse cube force that seems to assist gravity in keeping large mass accumulations stable. This force arises from an inverse square law potential. From the work above and in the gravitational context this force can be seen as arising from the Schr\"odinger equation $\nabla^2$ term's angular momentum part. It is as though this force gravitationally counters rotational {\it centrifugal\/} force. This facet seems to be a special aspect of the gravitational equilibrium for astrophysical mass accumulations.

*************

\section{Acknowledgements}
I am greatly indebted to Professors Clive Kilmister and Wolfgang Rindler for help, encouragement and inspiration over many years.

\end{document}